\documentclass[fleqn,usenatbib]{mnras}

\usepackage{newtxtext,newtxmath}

\usepackage[T1]{fontenc}

\DeclareRobustCommand{\VAN}[3]{#2}
\let\VANthebibliography\thebibliography
\def\thebibliography{\DeclareRobustCommand{\VAN}[3]{##3}\VANthebibliography}

\usepackage{graphicx}	
\usepackage{amsmath}	
\usepackage{dcolumn}
\usepackage{graphics}
\usepackage{slashed}
\usepackage{enumerate}
\usepackage{url}
\usepackage{color}
\usepackage{xcolor}
\usepackage{ulem}
\usepackage{comment}
\usepackage{tikz}

\definecolor{color1}{RGB}{191, 0, 255}

\title{Robustness of Magnetic Field Amplification in Neutron Star Mergers}

\author[R. Aguilera-Miret et al.]{
	Ricard Aguilera-Miret,$^{1,2}$\thanks{E-mail: ricard.aguilera.miret@uni-hamburg.de (RA-M)}
	Jan-Erik Christian,$^{1}$
	Stephan Rosswog,$^{1,3}$
	Carlos Palenzuela,$^{2,4,5}$
\\
$^{1}$University of Hamburg, Hamburger Sternwarte, Gojenbergsweg 112, 21029, Hamburg, Germany\\
$^{2}$Institute of Applied Computing \& Community Code (IAC3),  Universitat  de  les  Illes  Balears,  Palma  de  Mallorca, E-07122,  Spain\\
$^{3}$The Oskar Klein Centre, Department of Astronomy, AlbaNova, Stockholm University, SE-106 91 Stockholm, Sweden\\
$^{4}$Departament  de  F\'{\i}sica,  Universitat  de  les  Illes  Balears,  Palma  de  Mallorca, E-07122,  Spain\\
$^{5}$Institut d'Estudis Espacials de Catalunya (IEEC), 08034 Barcelona, Spain
}

\date{Accepted XXX. Received YYY; in original form ZZZ}

\pubyear{\the\year{}}

\begin{document}
\label{firstpage}
\pagerange{\pageref{firstpage}--\pageref{lastpage}}
\maketitle

\begin{abstract}
The dynamics of a binary neutron stars merger is governed by physics under the most extreme conditions, including strong spacetime curvature, ultra-high matter densities, luminous neutrino emission and the rapid amplification of the initial neutron star magnetic fields. Here we systematically explore how sensitive the magnetic field evolution is to the total mass of the merging binary, to the mass ratio of its components, the stellar spins and to the equation of state. For this purpose, we analyze 16 state-of-the-art GRMHD simulations that employ a subgrid-scale model to account for the unresolved small-scale turbulence.  We find that strong and rapid amplification of the magnetic field to volume-averaged values of $\sim 10^{16}$~G in the high-density regions is a very robust outcome of a neutron star merger and this result is only marginally impacted by either mass, mass ratio, spin or equation of state.
\end{abstract}

\begin{keywords}
MHD -- neutron star mergers -- magnetic fields -- turbulence
\end{keywords}



\section{Introduction}
The first gravitational wave (GW) detection of a neutron star merger (GW170817) ~\citep{LVC-BNS} was accompanied by a firework of electromagnetic radiation~\citep{LVC-MMA}. The complementarity of the received information allowed for a number of major leaps forward on many long-standing problems. It allowed, for example, to restrict the propagation speed of gravity to extremely close to the speed of light, with a deviation of at most 1 part in $10^{15}$~\citep{abbott17d}. By placing constraints on the tidal deformability of the neutron stars in the last inspiral stages \citep{abbott17b} these observations also ruled out several dense matter equation of state.

The observed "kilonova" \citep{Arcavi2017,Coulter2017,Pian2017,Smartt2017,tanvir17,evans17}, in turn, showed beyond reasonable doubt that neutron star mergers are indeed major sources of r-process elements~\citep{Kasen2017,rosswog18a,Metzger2019LRR} and thus observationally confirmed the results of earlier theoretical work on this topic \citep{lattimer74,symbalisty82,eichler89,rosswog99,freiburghaus99b}. Apart from confirming neutron star mergers as r-process sites, the early blue part of the kilonova and the identification of strontium lines also underlined that a fair fraction of the initially extremely neutron rich matter ($Y_e\sim 0.05$) had been transformed to values beyond the critical value of $\approx 0.25$ where the nucleosynthesis changes abruptly \citep{korobkin12a,lippuner15}, thereby underlining the importance of weak interaction/neutrino transport during the merger process.

Neutron star mergers had also for a long time been associated with short gamma ray bursts (sGRBs) \citep{eichler89,narayan92,piran04,meszaros06,nakar07,nakar20}. However, while this association based on various theoretical arguments was very plausible, it was only really confirmed by the observation of the sGRB accompanying GW170817 \citep{abbott17d}. While the observational confirmation that neutron star merger can launch short GRBs is an important milestone, we are still left with the long standing puzzle of how the necessary outflows with Lorentz factors of several hundred \citep[e.g.][]{nakar20} can be achieved. Although the detailed launch process is still not fully understood, it is generally accepted that magnetic fields play (at least) a very important role in it. Magnetic fields are further key in unbinding matter from the post-merger remnant torus which plausibly contributes the bulk of the neutron star merger ejecta \citep{metzger08a,beloborodov08,siegel17a,fernandez19,miller19}. Moreover, magnetic fields are can convert axions to potentially observable photons \citep{barack19,edwards20,harris20,fiorillo22}. The understanding of how magnetic fields evolve in a neutron star merger is therefore of uttermost importance for essentially all post-merger signatures of a binary neutron star coalescence.

A study of the merger process of two neutron stars and its aftermath requires careful numerical relativity simulations. For recent reviews of this multifaceted subject we refer to the literature, see \citep[e.g.][]{shibata16,Paschalidis2017,baiotti17,Duez2019,Shibata2019,perego19a,Ciolfi2020b,Palenzuela2020,bernuzzi20a,radice20,sarin21}. Many general relativistic magnetohydrodynamics (GRMHD) simulations reported magnetic field energies of $10^{51}$~ergs or even higher during a binary neutron star (BNS) merger phase. The initial magnetic fields of the merging neutron stars can be amplified by various processes, such as the winding effect due to the large-scale differential rotation~\citep{Duez2006b}, alpha-dynamo~\citep{kiuchi24}, turbulent dynamo~\citep{turbulent_dynamo},the Kelvin-Helmholtz instability (KHI)~\citep{price06,kiuchi15} and the magneto-rotational instability (MRI)~\citep{balbus91,balbus98,duez2006a,siegel13,kiuchi14}. 

The first instability to manifest in the binary system is the Kelvin–Helmholtz instability (KHI), which arises in the shear interface where the fluids from the two stars move in opposite directions. In this region, small perturbations are amplified, leading to the formation of vortexes across a range of spatial scales. The magnetic field is wounded up in these vortices and its magnitude grows on a time scale $\tau_{\rm KHI} \propto \lambda$, where $\lambda$ is the length scale of the initial perturbation. This means that perturbations of the shortest length scales that can be numerically resolved grow fastest. The non-linear interaction of these vortices leads to the development of turbulence. If the magnetic energy density is much weaker than the turbulent kinetic energy density, the turbulent flow can efficiently amplify the magnetic field by randomly stretching, twisting, and folding its field lines. This process, commonly known as turbulent (or small-scale) dynamo, also requires to capture numerically the very short length scales present in the system. Even the highest resolved simulations to date are likely still far from completely resolving the instabilities and turbulence associated to the MHD dynamics~\citep{kiuchi18,2023arXiv230615721K}.

Various approaches exist to account for the effects of unresolved MHD processes. One of them is the imposition of a very strong, purely poloidal, large-scale magnetic field $\gtrsim 10^{15}$ G just before the merger ~\citep[e.g.][]{ruiz16,kiuchi18,ciolfi2019,ciolfi2020collimated,ruiz2020,mosta2020,combi2023b,Most2023,2023arXiv230615721K}. Starting with such a magnetic field is qualitatively justified to compensate for the small scale dynamics that can not be captured with simulations of insufficient resolution. However, the magnetic energy will not be amplified up to the energy mentioned above when starting with a more realistic value of $\lesssim 10^{11}$~G (as expected in Gyr-old NSs~\citep{tauris17}) in low resolution simulations. A drawback of neglecting the small scale dynamics is the overemphasis of the initial magnetic fields influence. This can be attributed to the magnetic field keeping its memory, i.e. an imprint of its initial topology and strength, in its amplified state~\citep{aguilera2020,palenzuela22,kiuchi18,mosta2020,aguilera23,aguilera24}, even though the memory of a realistic initial magnetic field would be irreconcilably disturbed by the turbulent regime. It was shown in Ref.~\citet{aguilera22} that the initial magnetic field strength needs to be below $\sim10^{14}$~G for the turbulent regime to erase the memory of the initial field.

An alternative way to compensate for the lack of sufficient resolution involves the use of Large-Eddy Simulations (LESs). In this technique, one simulates the computationally cheaper large scales of the system and models the small ones by means of a subgrid-scale (SGS) model. A SGS model is a numerical approach used in computational fluid dynamics to approximate the effects of small, unresolved turbulent scales on the larger, resolved flow field. By doing so, and modifying the GRMHD evolution equations, one can capture part of the unresolved dynamics of the system~\citep{zhiyin15}. By improving the evolution equations with the gradient SGS model~\citep{leonard75,muller02a,grete16,grete17phd}, one can recover part of the effects induced by the unresolved SGS dynamics on the resolved scales. Combining the gradient SGS model, high-order numerical schemes and high-resolution simulations allow to improve the accuracy of the magnetic amplification in a turbulent regime, both in box~\citep{carrasco19,vigano20} and in full BNS merger simulations~\citep{aguilera2020,palenzuela22,aguilera22,aguilera23,aguilera24}. Using these techniques it was possible, for the first time, to achieve convergence of the volume-averaged magnetic field at the end of the strong turbulent phase induced by the KHI~\citep{palenzuela22}.

In this work we investigate the turbulent amplification of the magnetic field within the first $10$~ms after the merger in 16 GRMHD simulations with various initial conditions. Our earlier work~\citep{palenzuela22} shows that after this time the magnetic field decreases slightly to subsequently increase again due to magnetic winding. We utilize the same SGS based approach as in \citet{palenzuela22} to achieve convergent results. For the BNS system we vary the total masses, mass ratios, stellar spins and the equations of state (EoSs). We explicitly include some cases for the cold EoS that are ruled out by the tidal deformability of GW170817. This allows us to showcase the robustness of our results, where even under extreme conditions the maximum volume-averaged magnetic field in the turbulent phase reaches values of $\sim 10^{16}$~G in the higher-density region ($\rho > 10^{13}$ g cm$^{-3}$; "bulk") of the remnant, with local values reaching $\sim 10^{17}$~G. In the surrounding lower density "envelope" ($5\times 10^{11}$ g cm$^{-3} < \rho < 10^{13}$ g cm$^{-3}$) the volume-average magnetic field reaches values close to $\sim 10^{15}$~G.

Our paper is organized as follows. In Section ~\ref{sec:setup} we summarize the numerical setup, including the evolution equations, numerical techniques, the EoS, initial conditions and analysis quantities. In Section~\ref{sec:results} we present and analyze the numerical results. And finally, we summarize our results and present our conclusions in Section~\ref{sec:conclusions}.

\section{Setup} \label{sec:setup}

The concept and mathematical foundations behind LES with a gradient SGS approach have been extensively explained in previous works (and references therein) in the context of classical~\citep{vigano19b} and relativistic MHD~\citep{carrasco19,vigano20,aguilera2020,palenzuela22,aguilera22}, to which we refer for details and further references. In the present work, we vary the total mass, the mass ratio, the stellar spins and the cold EoS of a neutron star binary system. The corresponding initial models are then evolved in time with LESs of the full GRMHD equations supplemented by the the gradient SGS model. The evolution equations, numerical methods and setup are almost identical as those in~\citet{palenzuela22}, which we now summarize briefly for completeness.

\subsection{Evolution equations: GRMHD LES} 
\label{sec:equations}

The evolution of the spacetime geometry is determined by the Einstein equations. The covariant field equations can be written as a  hyperbolic evolution system by performing the established $3+1$ decomposition ~\citep[e.g.][]{bonabook,Palenzuela2020}. We use the covariant conformal Z4 formulation of the evolution equations~\citep{alic12,bezares17}. A summary of the final set of evolution equations for the spacetime fields, together with the gauge conditions setting the choice of coordinates, can be found in~\citep{palenzuela18}. We further use the common conventions of $G=c=1$ and a metric with signature $(-,+,+,+)$. The magnetized perfect fluid describing the star follows the GRMHD equations \citep[e.g.][]{shibatabook,palenzuela15}, which are written as a set of evolution equations for the conserved variables $ \left\lbrace D, S^i , U, B^i \right\rbrace$. These conserved fields are functions of the primitive fields, namely the rest-mass density $\rho$, the specific internal energy $\epsilon$, the velocity vector $v^{i}$ and the magnetic field $B^i$. The recovery of the primitive from the conserved fields requires first a closure relation for the pressure $p$ (i.e., the EoS) and the solution of a set of non-linear algebraic equations involving the Lorentz factor $W = (1-\gamma_{ij}v^iv^j)^{-1/2}$, with $\gamma_{ij}$ being the induced spatial metric. The full set of evolution equations, including all the gradient SGS terms, can be found in~\citet{vigano20} and \citet{aguilera2020}.

When applying the filtering to the GRMHD evolution system there appears new terms in the equations that are calculated using gradient SGS model. Each SGS term has associated a free  parameter of order unity, namely ${\cal C_M, C_N}$ and ${\cal C_T}$, which needs to be magnified to compensate for the numerical dissipation of the employed numerical scheme. Since we are mostly interested in the magnetic field dynamics, we follow our previous studies~\citep{carrasco19,vigano20,aguilera2020,palenzuela22,aguilera22,aguilera23,aguilera24,izquierdo24} and include only the SGS term appearing on the induction equation with the pre-coefficient ${\cal C_M}= 8$. This has been shown to reproduce the magnetic field amplification more accurately, even compared to very high-resolution simulations~\citep{vigano20,aguilera2020}. We remind the reader that these SGS terms, by construction, vanish in the infinite resolution limit.

\subsection{Numerical methods}

As in our previous works~\citep{aguilera2020,palenzuela22,aguilera22}, we use the code {\sc MHDuet}, generated by the platform {\sc Simflowny} \cite{arbona13,arbona18} and based on the {\sc SAMRAI} infrastructure \citep{hornung02,gunney16}, which provides the parallelization and the adaptive mesh refinement. It uses fourth-order-accurate operators for the spatial derivatives in the SGS terms and in the Einstein equations (the latter are supplemented with sixth-order Kreiss-Oliger dissipation); a high-resolution shock-capturing (HRSC) method for the fluid, based on the  Lax-Friedrich flux splitting formula \citep{shu98} and the fifth-order reconstruction method MP5 \citep{suresh97}; a fourth-order Runge-Kutta scheme with sufficiently small time step $\Delta t \leq 0.4~\Delta x$ (where $\Delta x$ is the grid spacing); and an efficient and accurate treatment of the refinement boundaries when sub-cycling in time~\citep{McCorquodale:2011,Mongwane:2015}. A complete assessment of the implemented numerical methods can be found in \citet{palenzuela18,vigano19}.

The binary is evolved in a cubic domain of size $\left[-1228,1228\right]$~km in each direction. The inspiral is fully covered by seven Fixed Mesh Refinement (FMR) levels. Each consists of a cube with twice the resolution of the next larger one. In addition, we use one Adaptive Mesh Refinement (AMR) level, covering the regions where the density is above $5 \times 10^{12}~\rm{g~cm^{-3}}$ and providing a uniform resolution throughout the shear layer. With this grid structure, we achieve a maximum resolution of $\Delta x_{\rm min} = 60$ m covering at least the most dense region of the remnant.

\subsection{Equation of state} 
\label{sec:EoS}

In our simulations we employ a tabulated version of the EoS at zero temperature which is augmented by an approximate thermal contribution to pressure and internal energy. While the largest contribution to the pressure is provided by the cold EoS, in a BNS merger the temperature cannot be neglected. By assuming an ideal gas we can calculate the thermal contribution to the pressure as $p_{\rm th} = (\Gamma_{\rm th} - 1) \rho \epsilon$, where $\Gamma_{\rm th}= 1.8$ is the adiabatic index. This approach is frequently used in numerical relativity simulations, \citep[e.g.][]{kiuchi18,palenzuela22,aguilera24}.

In this work we use six different EoSs. When varying mass, mass ratio and spin we employ a tabulated version of the piecewise polytrope~\citep{read09} fit to the well known APR4~\citep{APR4} EoS at zero-temperature, with a modification to prevent superluminal speeds~\citep{Endrizzi2016}. This EoS was chosen because it is widely used and can easily be compared to the results of other groups. Additionally, the merger remnants do not collapse into black holes during the time span of the simulations, so that the magnetic fields have sufficient time to grow.

However, in order to explore the full effect the EoS has on the magnetic field amplification, we also use a highly modifiable relativistic mean field (RMF) EoS \citep{PhysRev.98.783,Duerr56,Walecka74,ToddRutel:2005fa,Chen:2014sca} developed by  \citet{Hornick:2018kfi}, which we refer to as the Frankfurt-Barcelona (FB) EoS. As is typical for a RMF approach, the nucleon interactions are modeled via the exchange of mesons, in this case the $\rho$, $\omega$ and $\sigma$ mesons. Fitting the coupling parameters of the resulting Lagrangian to properties of nuclear matter determined by terrestrial experiments calibrates the approach. In the setup used by Hornick et al. the symmetry energy $J$, its slope $L$ and effective nucleon mass $m^{*}/m$ are varied at saturation density. Of these parameters $m^{*}/m$ has the greatest effect on EoS and its stiffness, i.e. its increase of pressure as a function of energy density \citep{Hornick:2018kfi,Christian:2023byx}. Therefore we fix the values $J = 32\,\mathrm{MeV}$ and $L = 60\,\mathrm{MeV}$, which gives us the largest possible range of $m^{*}/m = 0.55 - 0.75$ for this model.

At low densities these EoSs are well constrained and compatible with chiral effective field theory results \citep{Hornick:2018kfi}. However, small effective masses are known to yield stiff EoSs \citep{Boguta:1981px}, which are not compatible with the tidal deformability constraint from GW170817 \citep{Abbott:2018wiz}. In this parametrization the EoS becomes soft enough to be compatible with GW170817 at $m^{*}/m = 0.65$ \citep{Hornick:2018kfi,Christian:2023byx}.

We deliberately investigate the behavior of smaller $m^{*}/m$ values as well in order to get a full picture of the influence of the EoS on the amplification of the magnetic field, even though we do not expect to observe the results in reality. Employing this EoS setup allows us to test the sensitivity of the magnetic field amplification with respect to nuclear properties.

\subsection{Conversion to primitive variables} 
\label{sec:Conversion}

The conversion from evolved or conserved fields to primitive or physical ones is performed by using the robust procedure introduced in~\citet{kastaun20}, which yielded excellent results in our previous work~\citep{palenzuela22,aguilera22}. Following a common practice in GRMHD simulations, the surrounding regions of the neutron stars are filled with a relatively tenuous, low-density {\it atmosphere}, i.e. a threshold lower value on density, which is necessary to prevent the failure of the HRSC schemes usually employed to solve the MHD equations. To minimize unphysical states of the conserved variables outside the dense regions, produced by the numerical discretization errors of the evolved conserved fields, we set the atmosphere density at $6 \times 10^{4}~\rm{g~cm^{-3}}$ (i.e., one order of magnitude lower than in~\citet{palenzuela22}, and more than ten orders of magnitude lower than the maximum density values). In addition, we apply the SGS terms in the volume where the turbulent dynamo will be most active. This covers the bulk and most of the envelope of the formed remnant.

\subsection{Initial conditions}

The initial data are created either with the {\sc Lorene}~\citep{lorene} or {\sc FUKA}~\citep{fuka1,fuka2} packages. The former is used for simulations with equal mass and without rotation, while the latter is used in all other cases.

Each star initially has a purely poloidal dipolar magnetic field that is confined to its interior, calculated from a vector potential component $A_ {\phi} \propto R^2 {\rm max}(p - p_{cut},0)$, where $p_{cut}$ is a hundred times the pressure of the atmosphere, and $R$ is the distance to the axis perpendicular to the orbital plane passing through the center of each star. The maximum magnetic field (at the centers) is $5 \times 10^{11}$ G for our standard equal-mass binary with APR4 EoS. Since the magnetic field is setup in terms of the pressure, it will depend on the total mass, mass ratio and EoS, such that for some binaries it reaches  peak values of a few $10^{13}$ G, while maintaining average values below $10^{12}$ G. For all cases, these values are sufficiently low that they do not affect the remnant's evolution. Also, these field strenghts are several orders of magnitude lower than the large initial fields that are used in other simulations ~\citep[e.g.][]{kiuchi15,ruiz16,kiuchi18,ciolfi2019,ciolfi2020collimated,ruiz2020} and not too far from the upper range of the oldest known neutron stars (millisecond pulsars and low-mass X-ray binaries~\citep{bahramian23}). Nevertheless, and as a reminder from previous work~\citep{aguilera22}, this initial configuration was shown to be rapidly forgotten during the merger process, as long as the initial values are not too large ($B_0 \lesssim 10^{14}$ G) and the scheme and resolution are able to capture the turbulent amplification mechanisms. In that case, the initially small magnetic field acts only as a seed for the KHI and the turbulent dynamo.

The masses were chosen so that the merger remnants will likely remain stable during our simulations. This prevents us from simulating scenarios in which the magnetic field is prevented from being amplified in the bulk region of the remnant because of a prompt collapse. For this reason we limit the total mass to about $2.7\,M_\odot$ for all our simulations. Although this is a rather conservative estimate of the threshold total mass \citep{Kochankovski:2025lqc}, such a choice does not affect our results significantly, as will be shown in section~\ref{sec:results}. On the lower end of the mass range, we simulate a merger where both stars only have $0.9\,M_\odot$. This is an extreme low mass system, significantly below what would be predicted by core-collapse supernovae simulations \citep{Tauris:2019sho,Muller:2024aod}. We then simulate two equal mass systems where the total mass of the system lies between these extremes ($M_{\rm ADM}=\{2.26, 2.46\}$~$M_\odot$), as well as two systems with unequal masses ($M_{\rm ADM}=\{2.20, 2.64\}$~$M_\odot$). For these variations of mass we consider exclusively the non rotational stars.

Conversely, when we vary the initial spin periods of the neutron stars, the mass is kept identical for all simulations. The slowest spinning star has a period of $50$ ms, since an even larger value would be nearly indistinguishable from the non-rotating case given that the simulation starts $\sim 15$~ms before the merging point. We perform five simulations with various spin periods of $\{0.7, 10, 15, 50\}$~ms. This is done to probe whether by maximizing the kinetic energy the contribution of the KHI to the magnetic field amplification is increased significantly.

All simulations described above use the APR4 EoS for the description of matter. However, to investigate the effect of the EoS we run five additional simulations with the FB EoS described in section~\ref{sec:EoS}. The total mass of the system is kept at the same value of $M_{\rm ADM}=2.68\,M_\odot$ as before, and no spinning cases are considered. These FB EoSs range from implausibly stiff with the {\ttfamily FB55} case to unrealistically soft with the {\ttfamily FB75} case. An argument could be made that only considering variations within a RMF model introduces some bias; however the APR4 EoS used for all other simulations does not produce meaningfully different results either, as will be shown in section~\ref{sec:results}.

For all simulations described previously only one initial condition was varied, to give us a better understanding of its effect on the magnetic field amplification. Nevertheless, we also include one more realistic scenario where the neutron stars are spinning and their masses are unequal, where the mass ratio $q=M_1/M_2=1.25$  and $M_{\rm ADM}= 2.7\,M_\odot$. All the characteristics of the simulations presented here are summarized in Table~\ref{tab:models}, together with some results that will be discussed in the following sections.

\begin{table*}
\centering
\caption{\textit{Parameters of the simulations:} Different simulations, with the total mass of the system, the mass ratio {\bf $q$}, the radii, the stellar spins, the volume-average of the magnetic field at $10$~ms after the merger both in the bulk and the envelope, the magnetic energy at $10$~ms after the merger in the bulk and the number of orbits before the merger. All systems start with a counterclockwise rotation. We use the following naming convention: simulations with our standard EoS and no stellar spins and equal masses, where we want to probe the dependence on the total mass, are labeled MX.XX. In similar cases where the mass ratio differs from unity, we add a "q" to the name. Cases where we explore the spin dependence are labeled SPXX, where the XX refers to stellar spin period. In the case SPr both the spins and the masses differ. Cases where we explore the sensitivity to the effective mass in the Frankfurt-Barcelona EoSs are labeled as FBXX, where XX refers to the effective mass. MRLES refers to the convergent simulation in~\citet{palenzuela22}.}
\begin{tabular}{|c|cccccccccc|}
\hline
\textbf{Name} & \multicolumn{1}{c|}{\textbf{\begin{tabular}[c]{@{}c@{}}M$_{ADM}$\\ {[}M$_\odot${]}\end{tabular}}} & \multicolumn{1}{c|}{\textbf{q}} & \multicolumn{1}{c|}{\textbf{\begin{tabular}[c]{@{}c@{}}Radii$_1$ [km]\end{tabular}}} & \multicolumn{1}{c|}{\textbf{\begin{tabular}[c]{@{}c@{}}Radii$_2$ [km]\end{tabular}}} & \multicolumn{1}{c|}{\textbf{\begin{tabular}[c]{@{}c@{}}Spin$_1$ [ms]\end{tabular}}} & \multicolumn{1}{c|}{\textbf{\begin{tabular}[c]{@{}c@{}}Spin$_2$ [ms]\end{tabular}}} & \multicolumn{1}{c|}{\textbf{\begin{tabular}[c]{@{}c@{}}$|$B$|_{13}$ [G]\\ ($10$~ms)\end{tabular}}} & \multicolumn{1}{c|}{\textbf{\begin{tabular}[c]{@{}c@{}}$|$B$|_{10}^{13}$ [G]\\ ($10$~ms)\end{tabular}}} & \multicolumn{1}{c|}{\textbf{\begin{tabular}[c]{@{}c@{}}$|$E$|_{\rm mag}$ [erg]\\ ($10$~ms)\end{tabular}}} & \multicolumn{1}{c|}{\textbf{\begin{tabular}[c]{@{}c@{}}\# orbits\\\end{tabular}}} \\ \hline
\textbf{M1.80} & 1.80 & 1 & 10.22 & 10.22 & 0 & 0 & $7 \cdot 10^{15}$ & $2 \cdot 10^{14}$ & $6.0 \cdot 10^{49}$ & 9 \\ \hline
\textbf{M2.26} & 2.26 & 1 & 9.77 & 9.77 & 0 & 0 & $7 \cdot 10^{15}$ & $4 \cdot 10^{14}$ & $4.9 \cdot 10^{49}$ & 8 \\ \hline
\textbf{M2.46} & 2.46 & 1 & 9.56 & 9.56 & 0 & 0 & $4 \cdot 10^{15}$ & $4 \cdot 10^{14}$ & $1.1 \cdot 10^{49}$ & 7 \\ \hline
\textbf{M2.20q} & 2.20 & 1.2 & 9.98 & 9.64 & 0 & 0 & $2 \cdot 10^{15}$ & $1 \cdot 10^{14}$ & $2.4 \cdot 10^{49}$ & 8 \\ \hline
\textbf{M2.64q} & 2.64 & 1.2 & 8.76 & 9.06 & 0 & 0 & $7 \cdot 10^{15}$ & $3 \cdot 10^{14}$ & $1.4 \cdot 10^{50}$ & 5 \\ \hline
\textbf{SP0.7} & 2.70 & 1 & 8.84 & 8.84 & 0.7 & 0.7 & $2 \cdot 10^{15}$ & $4 \cdot 10^{14}$ & $6.6 \cdot 10^{48}$ & 5 \\ \hline
\textbf{SP10} & 2.70 & 1 & 8.84 & 8.84 & 10 & 10 & $2 \cdot 10^{15}$ & $5 \cdot 10^{14}$ & $5.2 \cdot 10^{48}$ & 5 \\ \hline
\textbf{SP15} & 2.70 & 1 & 8.84 & 8.84 & 15 & 15 & $2 \cdot 10^{15}$ & $7 \cdot 10^{14}$ & $5.2 \cdot 10^{48}$ & 5 \\ \hline
\textbf{SP50} & 2.70 & 1 & 8.84 & 8.84 & 50 & 50 & $2 \cdot 10^{15}$ & $7 \cdot 10^{14}$ & $4.9 \cdot 10^{48}$ & 5 \\ \hline
\textbf{SPr} & 2.70 & 1.25 & 8.55 & 9.14 & 1 & 10 & $3 \cdot 10^{15}$ & $6 \cdot 10^{14}$ & $ 1.8 \cdot 10^{49}$ & 5 \\ \hline
\textbf{FB55} & 2.70 & 1 & 10.17 & 10.17 & 0 & 0 & $9 \cdot 10^{15}$ & $6 \cdot 10^{14}$ & $1.5 \cdot 10^{50}$ & 5 \\ \hline
\textbf{FB60} & 2.70 & 1 & 9.58 & 9.58 & 0 & 0 & $8 \cdot 10^{15}$ & $7 \cdot 10^{14}$ & $1.4 \cdot 10^{50}$ & 4 \\ \hline
\textbf{FB65} & 2.70 & 1 & 9.13 & 9.13 & 0 & 0 & $7 \cdot 10^{15}$ & $7 \cdot 10^{14}$ & $9.8 \cdot 10^{49}$ & 4 \\ \hline
\textbf{FB70} & 2.70 & 1 & 8.54 & 8.54 & 0 & 0 & $8 \cdot 10^{15}$ & $7 \cdot 10^{14}$ & $1.1 \cdot 10^{50}$ & 4 \\ \hline
\textbf{FB75} & 2.70 & 1 & 7.67 & 7.67 & 0 & 0 & $6 \cdot 10^{15}$ & $5 \cdot 10^{14}$ & $4.5 \cdot 10^{49}$ & 4 \\ \hline
\textbf{MRLES} & 2.70 & 1 & 8.31 & 8.31 & 0 & 0 & $9 \cdot 10^{15}$ & $6 \cdot 10^{14}$ & $8.6 \cdot 10^{49}$ & 5 \\ \hline
\end{tabular}
\label{tab:models}
\end{table*}

\subsection{Analysis quantities}

We use several integral quantities to monitor the dynamics in different regions, like the averages of the magnetic field strength, the fluid angular velocity $\Omega \equiv \frac{d\phi}{dt} = \frac{u^{\phi}}{u^t}$ (where $u_a \equiv W (-\alpha, v_i )$ denotes the fluid four-velocity), with $\alpha$ being the lapse function, and the plasma beta parameter, $\beta = \frac{2 P}{B^2}$. The averages for a given quantity $q$ over a certain region ${\cal N}$ will be denoted generically by:
\begin{eqnarray}
	{<}q{>}_a^b = \frac{\int_{{\cal N}} q \, d{\cal N}}{\int_{{\cal N}} d{\cal N}} ,
\end{eqnarray}
where ${\cal N}$ stands for a volume $V$, a surface $S$, or a line $\ell$, and the integration is restricted to regions where the mass density is within the range $(10^a, 10^b)\, {\rm g/cm^3}$. If $b$ is omitted, it means no upper density cut is applied. From here on we will denote 
\begin{itemize}
	\item the \textbf{bulk} as the densest region of the remnant with densities $\rho > 10^{13}~{\rm g~cm}^{-3}$, and 
	\item the \textbf{envelope}  as the region satisfying $5 \times 10^{10} {\rm g~cm}^{-3} < \rho < 10^{13} ~{\rm g~cm}^{-3}$. 
\end{itemize}
In particular, we define averages over the bulk of the remnant as ${<}q{>}_{13}$, and averages over the envelope as ${<}q{>}^{13}_{10}$. We also compute global quantities, integrated over the whole computational domain, such as the total magnetic energy $E_{\rm mag}$, kinetic energy $E_{\rm kin}$, thermal energy $E_{\rm th}$ and rotational kinetic energy $E_{\rm rot}$, as given by:
\begin{eqnarray}
	E_{\rm mag} &=& \frac{1}{2}\int B^2 \sqrt{\gamma} dx^3 ,    
	\\
	E_{\rm kin} &=& \frac{1}{2}\int \rho v^2 \sqrt{\gamma} dx^3 ,    
	\\
	E_{\rm th} &=& \int \rho W (\epsilon - \epsilon_{\rm cold}) \sqrt{\gamma} dx^3 ,
	\\
	E_{\rm rot} &=& \frac{1}{2}\int \Omega T^t_{\phi} \sqrt{\gamma} dx^3 ,
\end{eqnarray}
where $T^t_{\phi}$ are the time-azimuthal components of the stress-energy tensor for a perfect fluid~\citep{vigano20}, $\Omega = d\phi/dt = u^\phi/u^t$ is the angular fluid velocity, and $\sqrt{\gamma}$ is the square root of the determinant of the spatial part of the metric~\citep{shibatabook}. In addition, we compute the spectral distribution of the kinetic and magnetic energies over the spatial scales. For the magnetic spectra, we also calculate the poloidal and toroidal contributions separately. Further details of the numerical procedure to calculate the spectra can be found in~\citet{aguilera2020,vigano19,vigano20} and \citet{palenzuela22}. With these energy spectra we can define the energy-weighted average wave-number $\langle k \rangle$ and its associated length scale $\langle L \rangle$, which can be calculated using
\begin{equation}
	\langle k \rangle \equiv \frac{\int_k k\,{\cal E}(k) \,dk} {\int_k {\cal E}(k)\, dk} \Rightarrow
	\langle L \rangle = \frac{2\pi}{\langle k \rangle}  ~,
\end{equation}
This length scale represents the typical \textit{coherent} scale of the structures present in the field.

\section{Results} \label{sec:results}

\begin{figure*}
	\centering
	\includegraphics[width=0.282\linewidth, clip=true, trim=1cm 4cm 1cm 1cm]{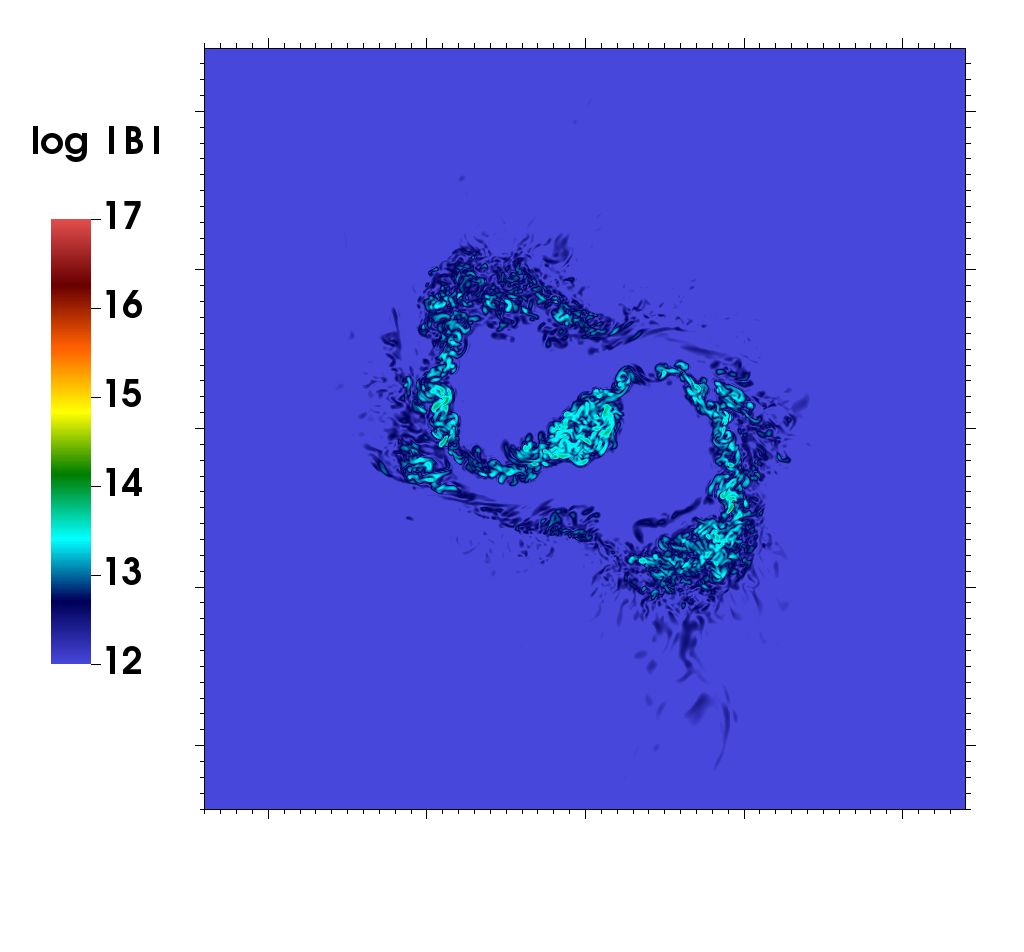}
	\includegraphics[width=0.233\linewidth, clip=true, trim=7cm 4cm 1cm 1cm]{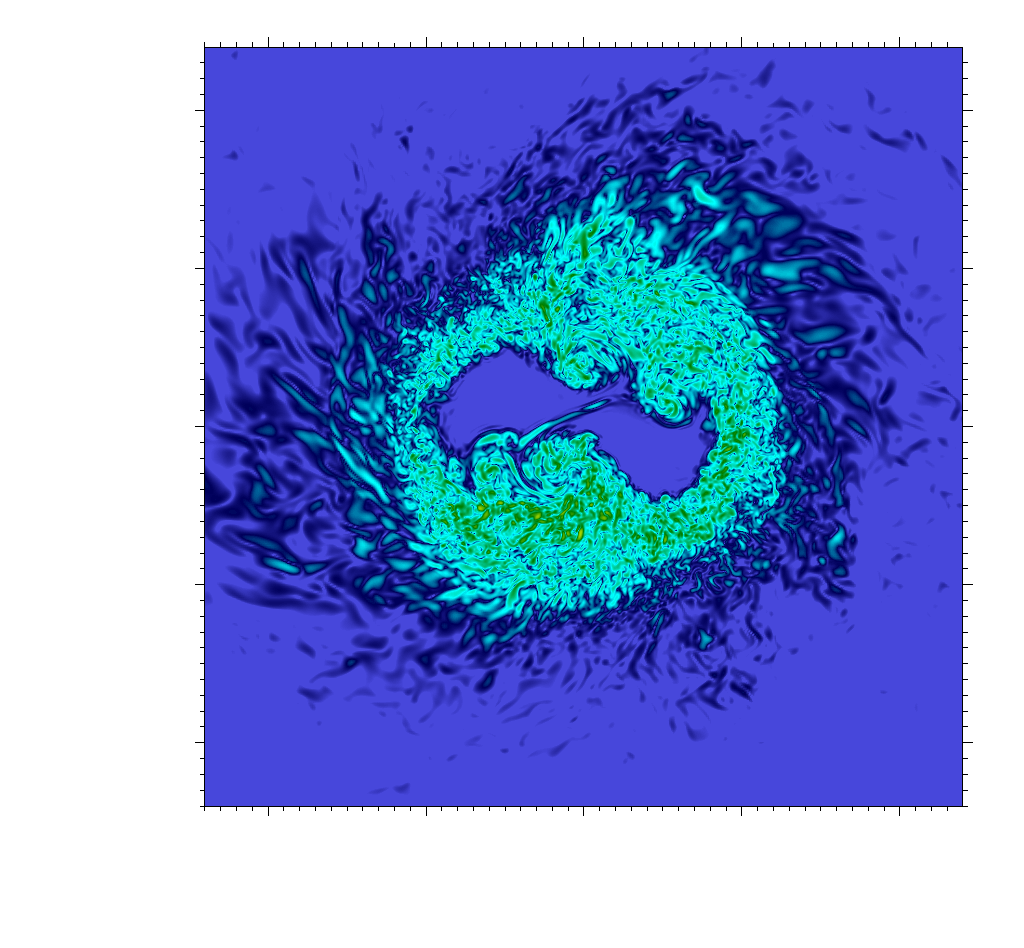}
	\includegraphics[width=0.233\linewidth, trim=7cm 4cm 1cm 1cm, clip]{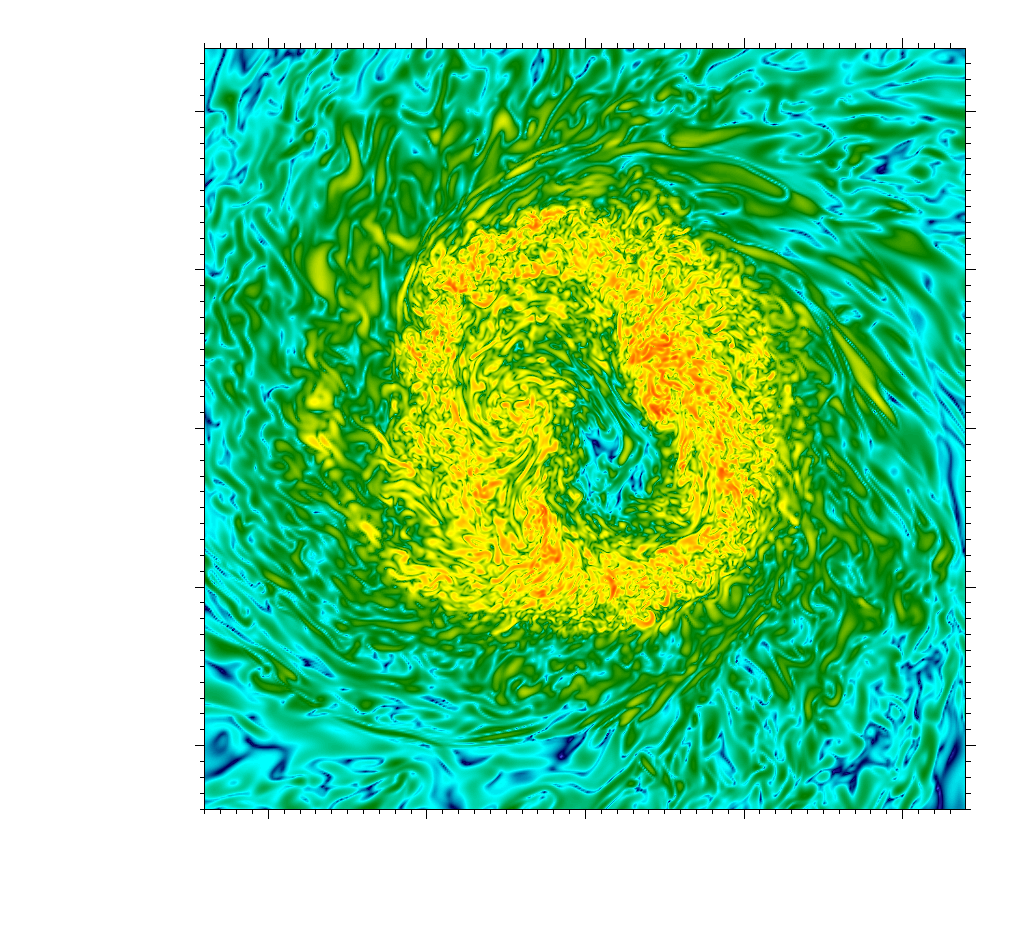}
	\includegraphics[width=0.233\linewidth, trim=7cm 4cm 1cm 1cm, clip]{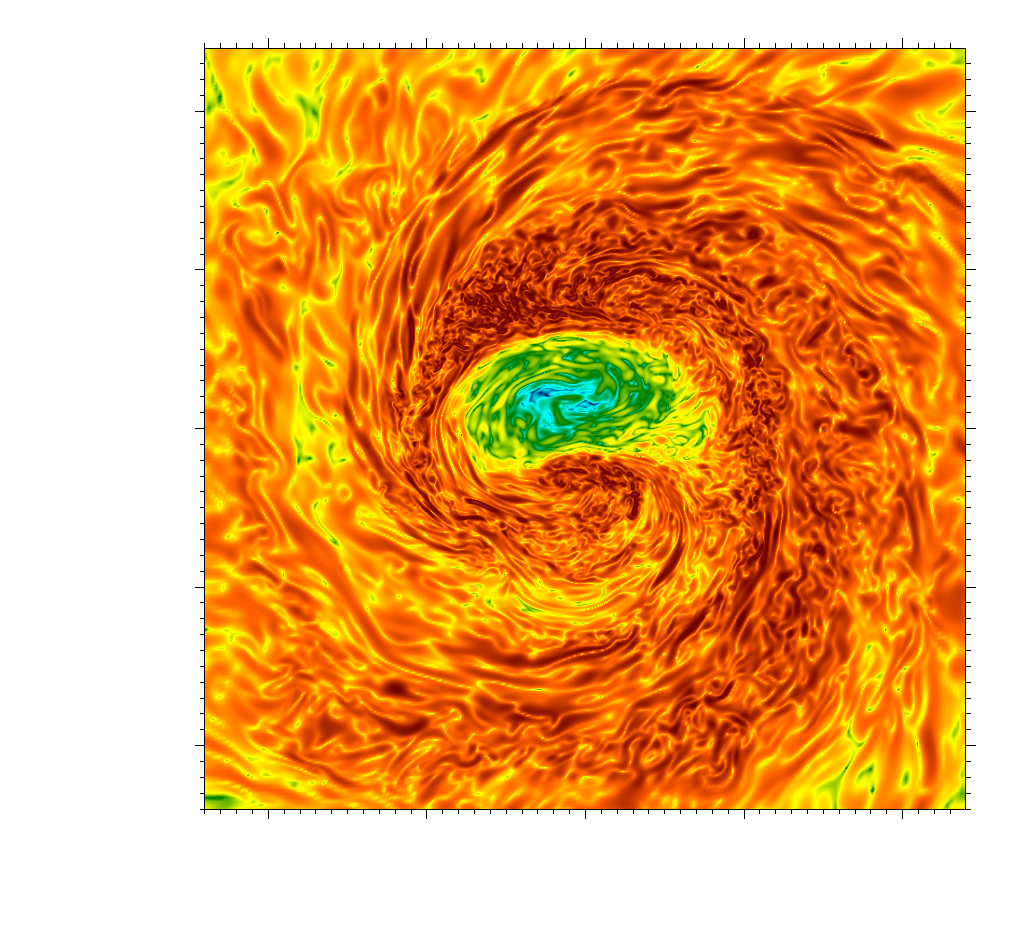}\\
	\includegraphics[width=0.282\linewidth, trim={1cm 4cm 1cm 1cm}, clip]{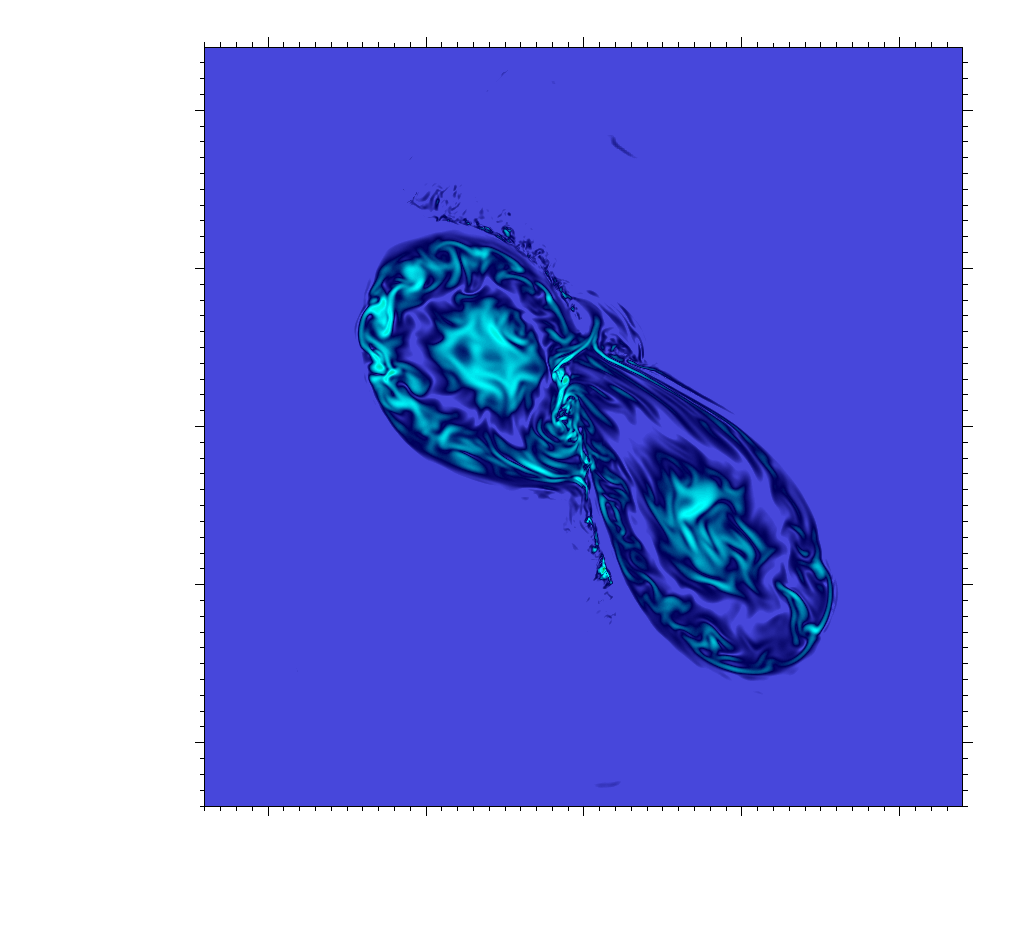}
	\includegraphics[width=0.233\linewidth, trim={7cm 4cm 1cm 1cm}, clip]{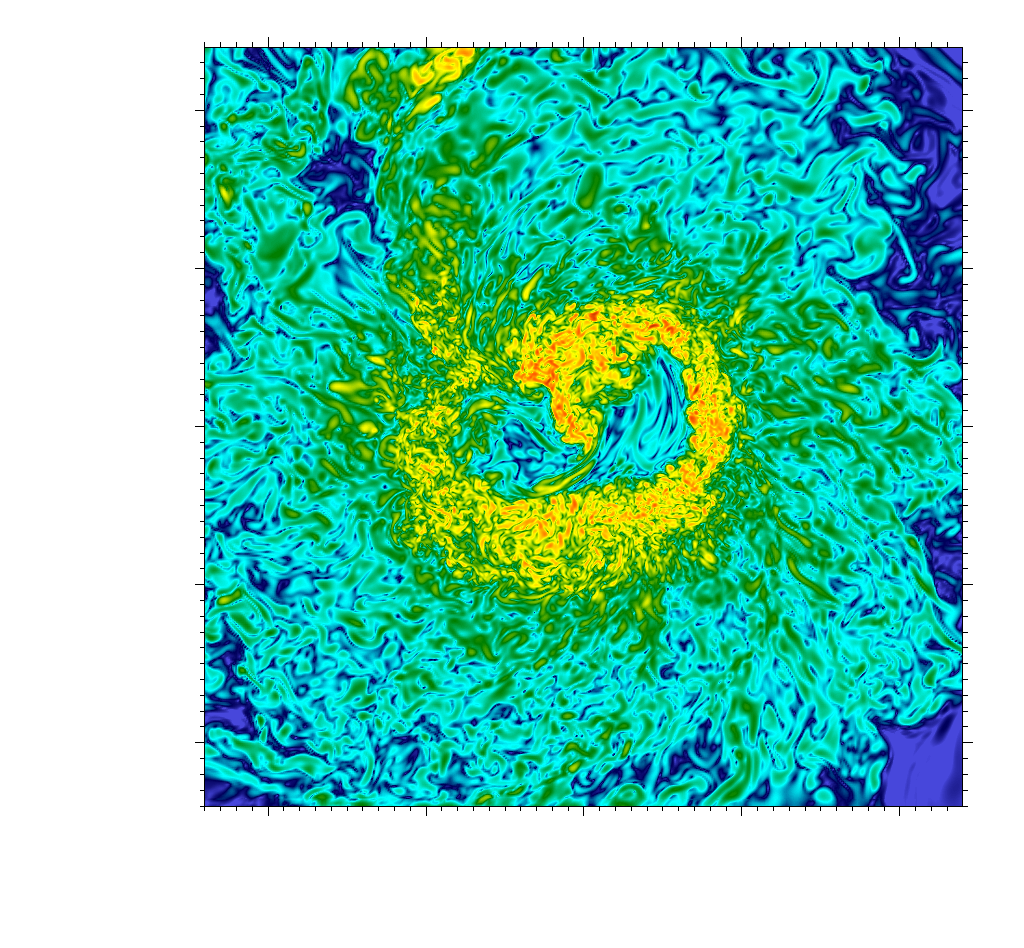}
	\includegraphics[width=0.233\linewidth, trim={7cm 4cm 1cm 1cm}, clip]{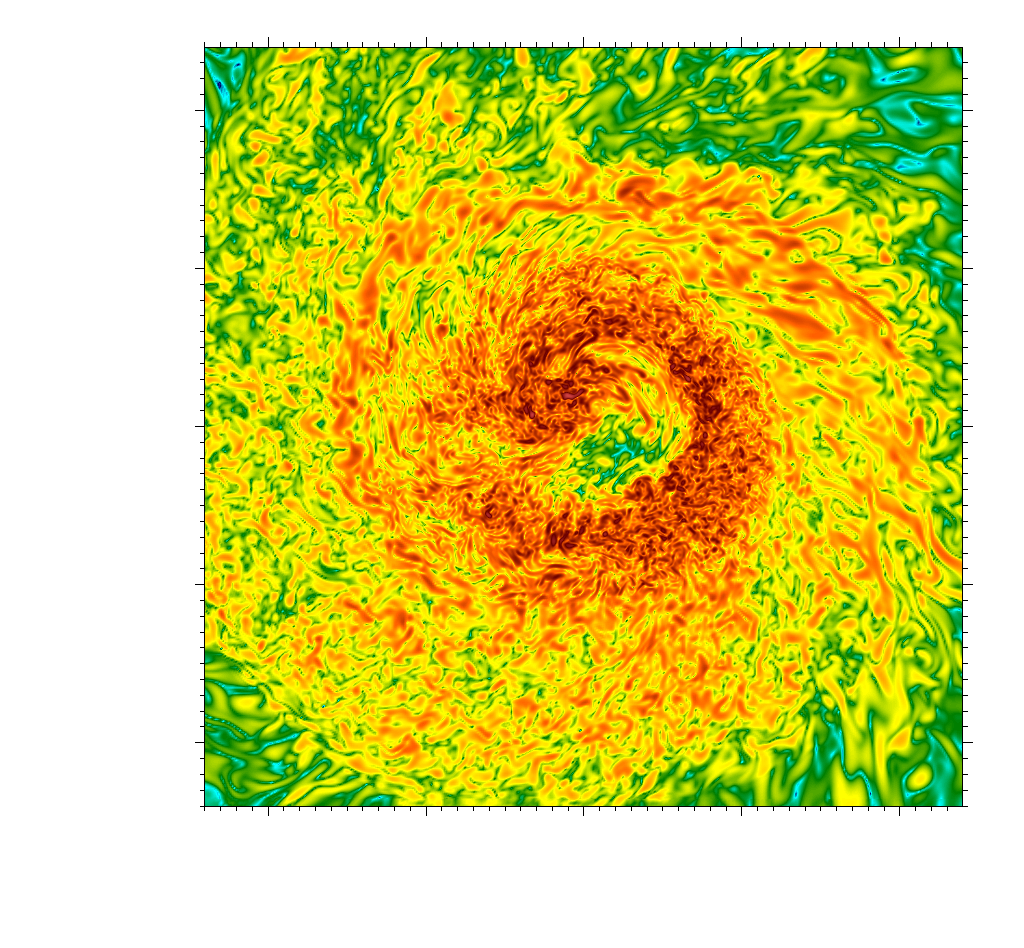}
	\includegraphics[width=0.233\linewidth, trim={7cm 4cm 1cm 1cm}, clip]{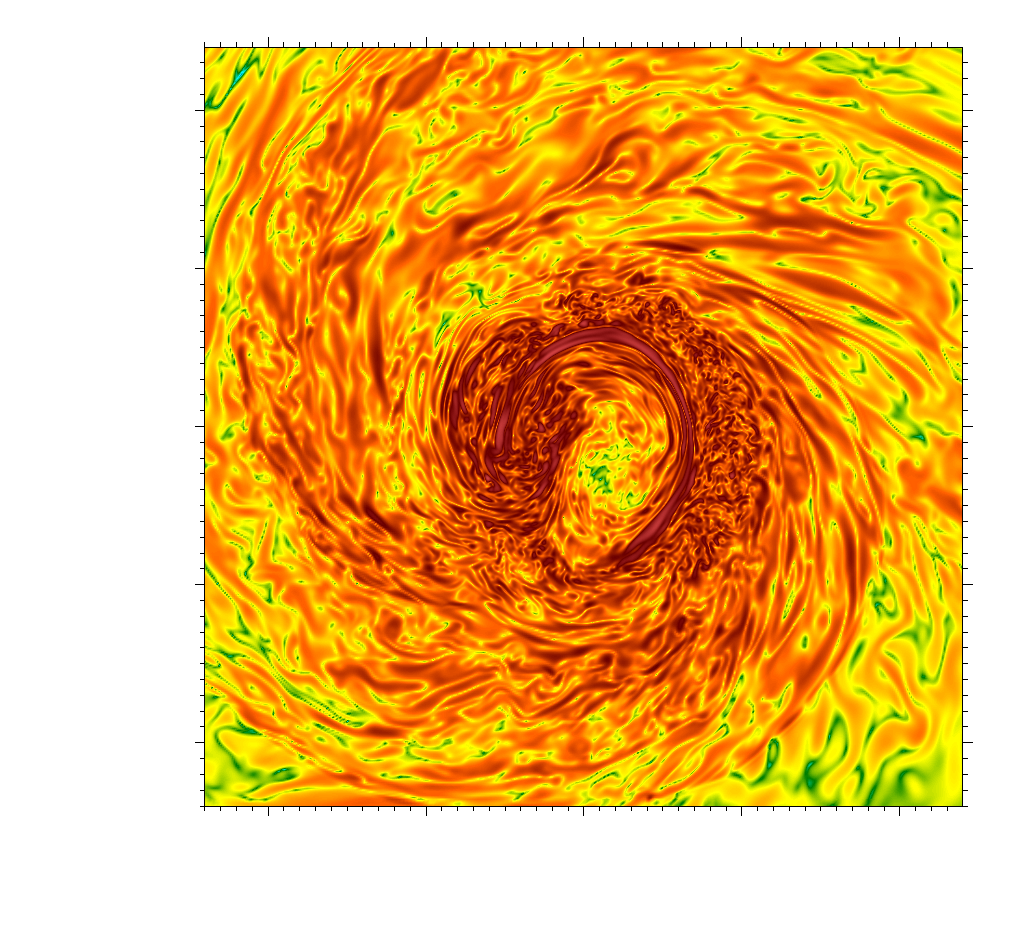}\\
	\includegraphics[width=0.282\linewidth, trim={1cm 2cm 1cm 1cm}, clip]{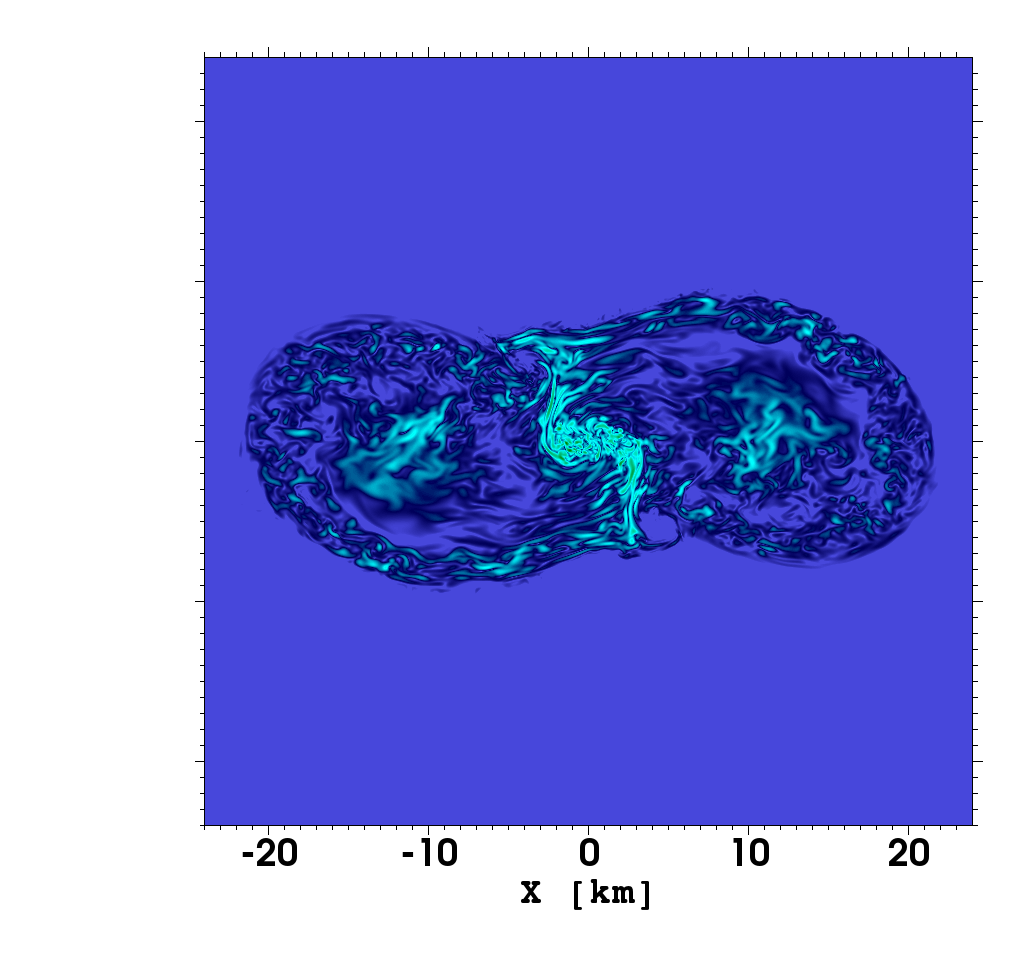}
	\includegraphics[width=0.233\linewidth, trim={7cm 2cm 1cm 1cm}, clip]{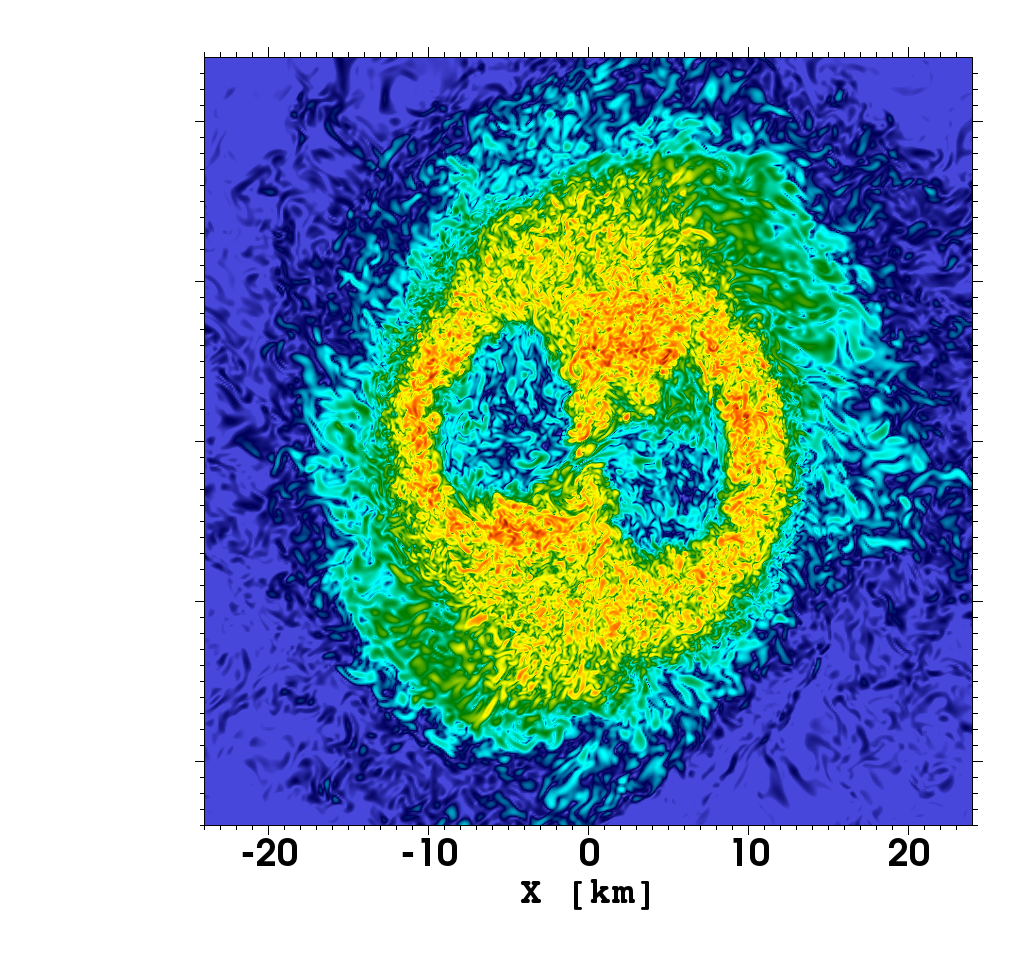}
	\includegraphics[width=0.233\linewidth, trim={7cm 2cm 1cm 1cm}, clip]{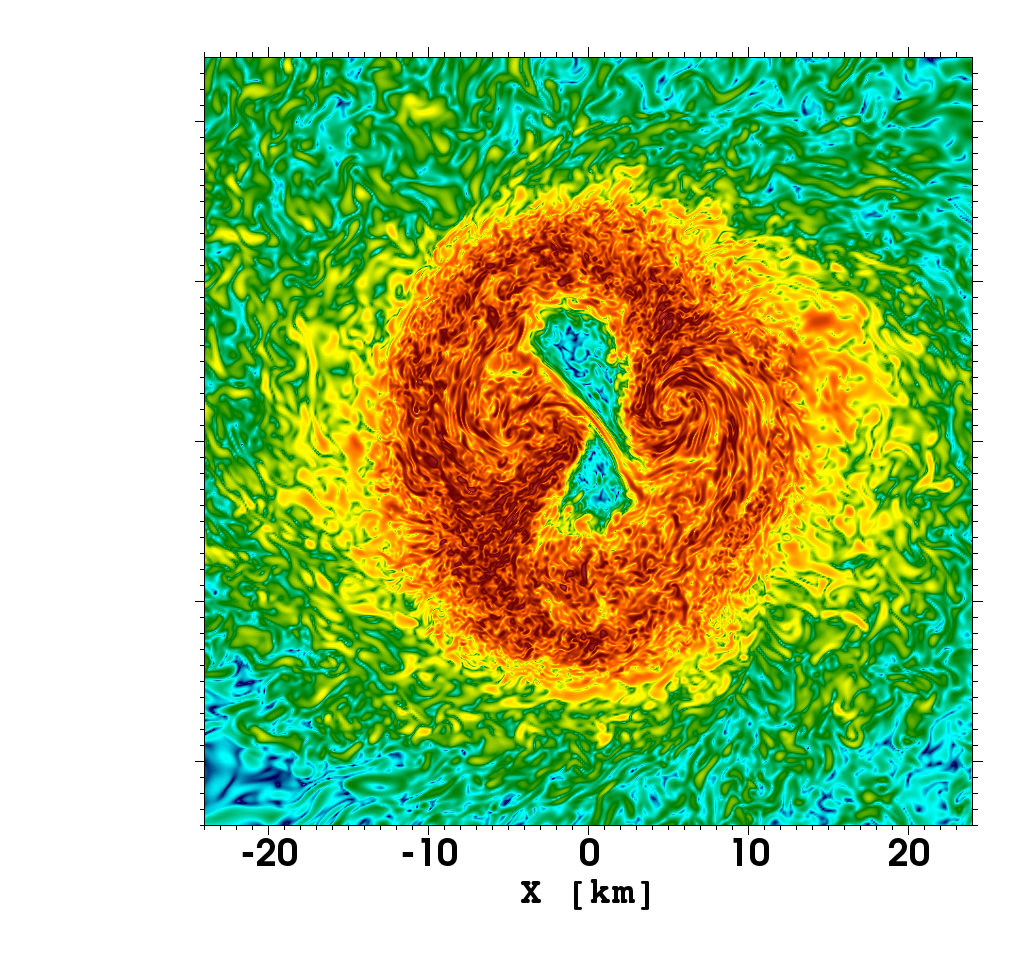}
	\includegraphics[width=0.233\linewidth, trim={7cm 2cm 1cm 1cm}, clip]{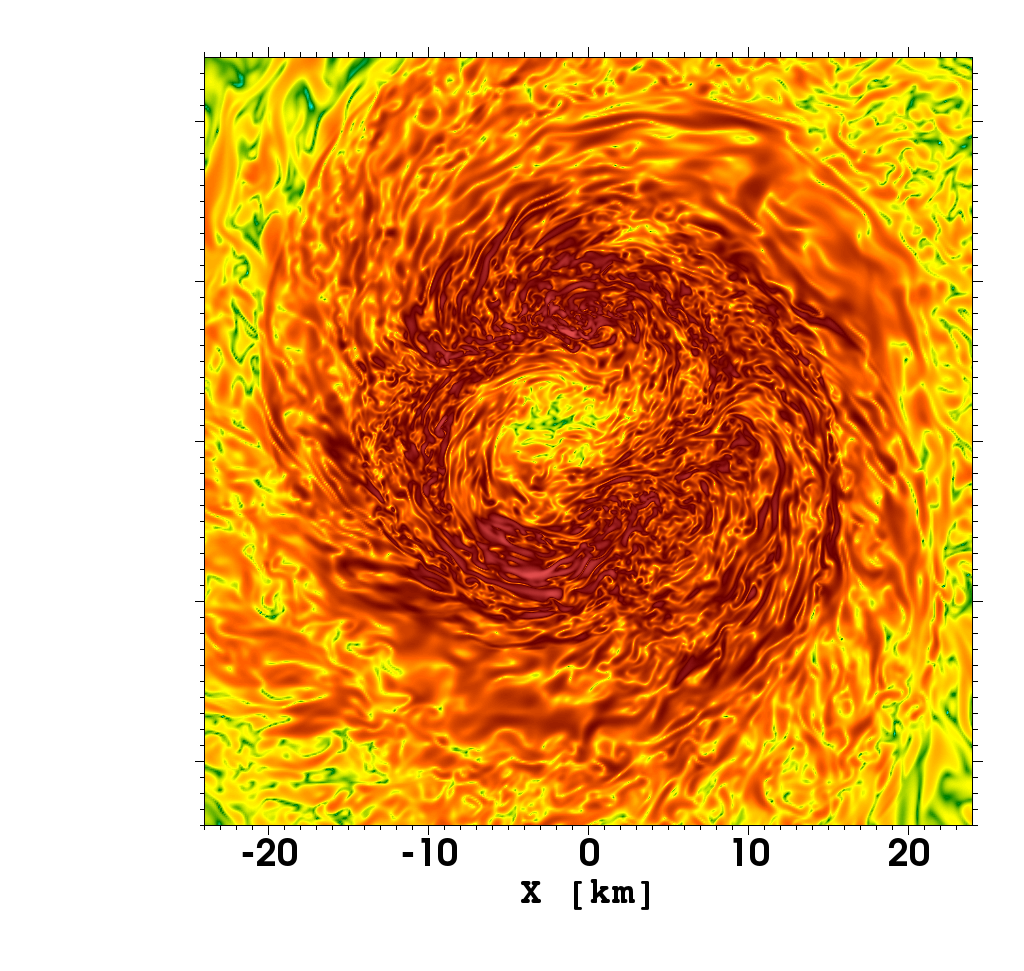}\\
	\caption{\textit{Magnetic field intensity}. Slices of the {\ttfamily M2.46} ($M_{ADM} = 2.46$, $q = 1$, spinless, EoS $=$ APR4), {\ttfamily SPr} ($M_{ADM} = 2.7$, $q = 1.25$, Spin$_1=1$~ms, Spin$_2 = 10$~ms, EoS$=$ APR4) and {\ttfamily FB65} ($M_{ADM} = 2.7$, $q = 1$, spinless, EoS $= m/m^* = 0.65$) simulations (first, second and third rows, respectively) at times $\{0.25, 2.5, 5, 10\}$~ms after the merger (in first, second, third and fourth columns respectively). The colourscale represents the magnetic field intensity.}
	\label{fig:orbital_plane}
\end{figure*}

\begin{figure*}
	\centering
	\includegraphics[width=0.45\linewidth, trim={0 2cm 2cm 2cm}, clip]{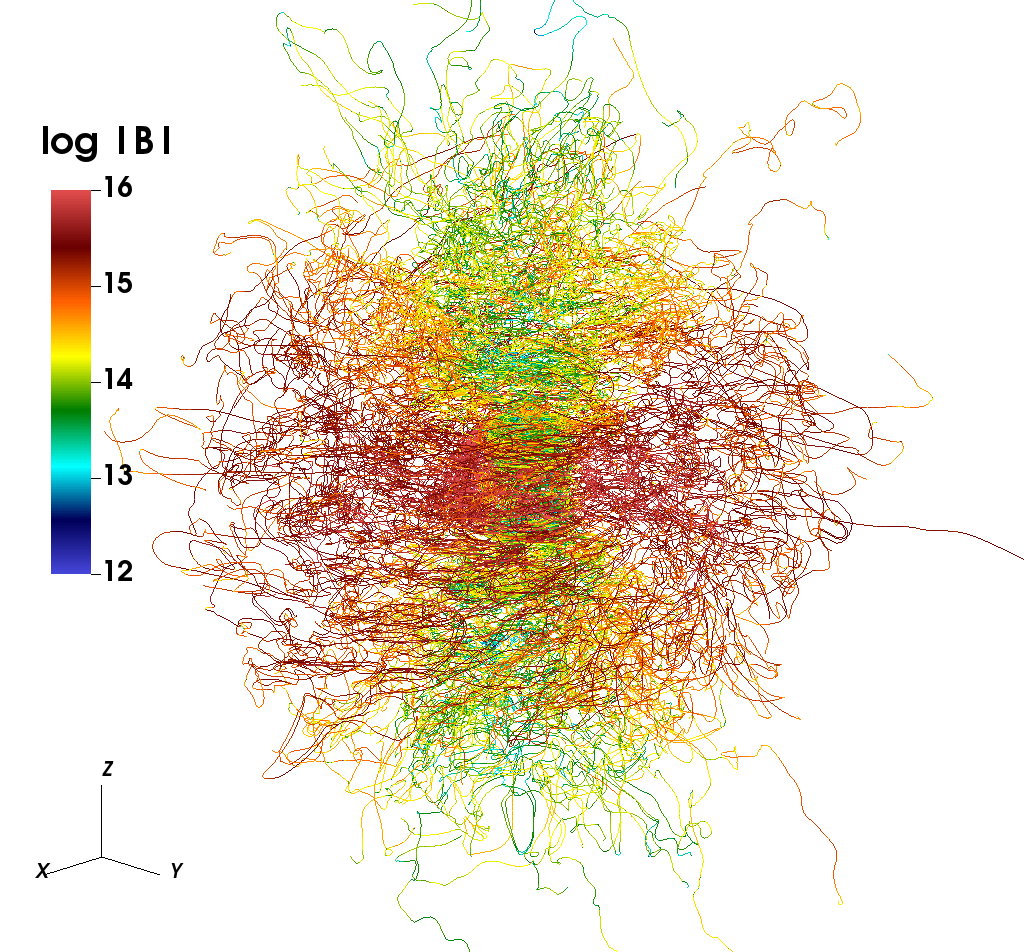}
	\includegraphics[width=0.45\linewidth, trim={0 2cm 2cm 2cm}, clip]{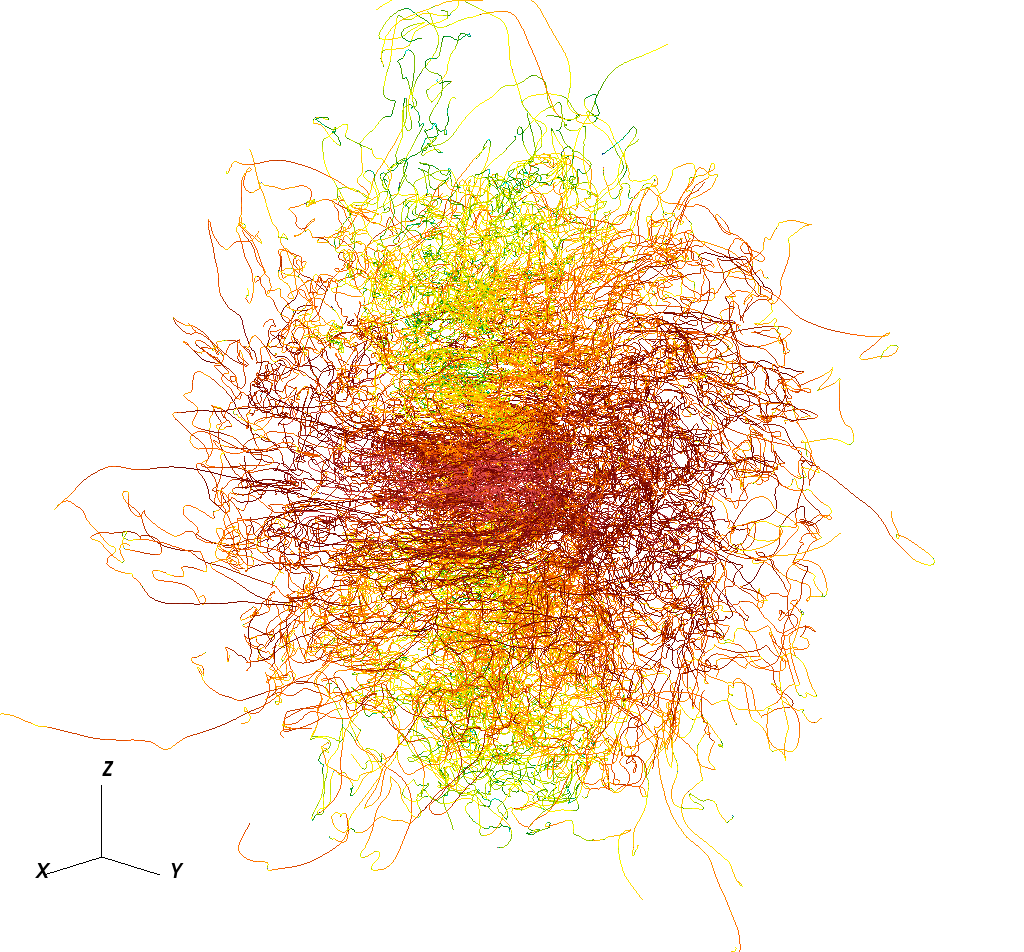}\\
	\includegraphics[width=0.45\linewidth, trim={0 2cm 2cm 2cm}, clip]{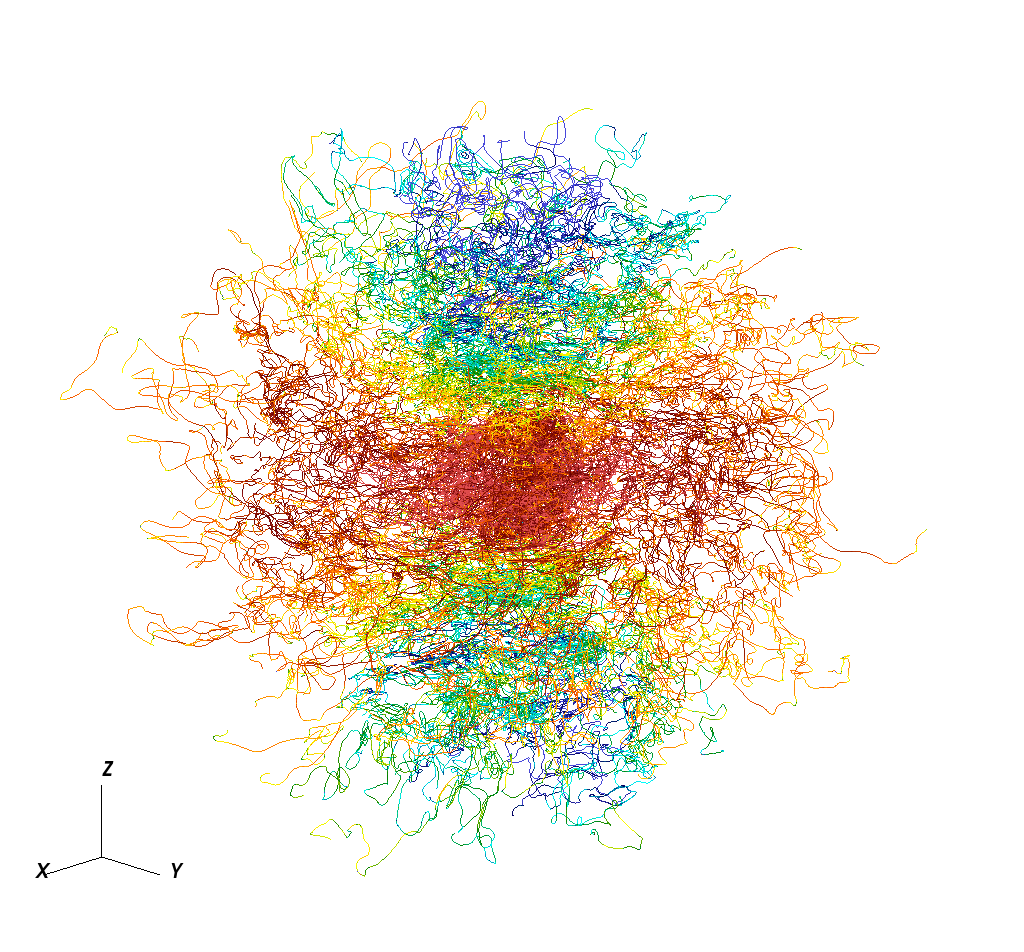}
	\caption{\textit{Magnetic streamlines}. Streamlines of the magnetic field, at $t = 10$~ms after the merger. From left to right and from top to bottom, {\ttfamily M2.46} ($M_{ADM} = 2.46$, $q = 1$, spinningless, EoS$=$ APR4), {\ttfamily SPr} ($M_{ADM} = 2.7$, $q = 1.25$, Spin$_1=1$~ms, Spin$_2 = 10$~ms, EoS$=$ APR4) and {\ttfamily FB65} ($M_{ADM} = 2.7$, $q = 1$, spinningless, EoS$=$ $m/m^* = 0.65$). The panels employ seeds spherically distributed isotropically for the streamline integration.}
	\label{fig:3D_streamlines}
\end{figure*}

In this section we will show the analysis of all the performed simulations. We also include the convergent {\ttfamily MRLES} results from ~\citet{palenzuela22} for comparative purposes. In this reference, we compared the results of a $\Delta x=60$~m resolution simulation including the gradient SGS model ({\ttfamily MRLES}) to a $\Delta x=30$~m resolution simulation activating or deactivating the gradient SGS model, {\ttfamily HRLES} and {\ttfamily HR}, respectively. We obtained a convergent result in all three simulations for the amplification of the volume-average magnetic field in the bulk region of the remnant and comparable results in the envelope.

Since we are only interested in the turbulent amplification of the magnetic field, we stop our simulations at $10$ ms after the merger. The exponential growth of the magnetic field due to the KHI and the turbulent dynamo only takes place roughly during this time interval. Note that we only show the results for three illustrative cases of the simulations we performed (namely {\ttfamily M2.46}, {\ttfamily SPr} and {\ttfamily FB65}) when plotting slices or magnetic field streamlines. The choice of these particular examples is rather arbitrary, as any other simulation within that set would represent their cases equally well.

\subsection{Qualitative dynamics}

All simulations are performed starting from the same coordinate separation (i.e., $45$~km). The simulations with less massive neutron stars take longer to merge, with the M1.80 case taking as long as nine orbits.

In Fig.~\ref{fig:orbital_plane} we present different slices of the orbital plane of the {\ttfamily M2.46}, {\ttfamily SPr} and {\ttfamily FB65} simulations (first, second and third column, respectively) at times $\{0.25, 2.5, 5, 10\}$~ms after the merger. The color scale represents the magnetic field intensity. At the beginning of the simulations, the magnetic field intensity is very low. Just after the merger, some eddies appear in the interface layer between the neutron stars. Since here the stellar fluids shear against each other, this interface becomes Kelvin-Helmholtz unstable with the shortest resolvable wavelengths growing fastest in the form of eddies. Later on, the remnant starts to bounce and turbulence develops. As the simulations evolve, the remnant keeps bouncing, thus triggering further turbulence, and the magnetic field is amplified up to $10^{17}$~G in some regions at the end of the timespan of the simulations, i.e. $10$~ms after the merger, where the structure of the magnetic field behaves with only minor differences between all the simulations.

In Fig.~\ref{fig:3D_streamlines} we show the magnetic streamlines of the three simulations at $t = 10$~ms after the merger. In the bulk, we find the strongest intensity of the magnetic field. In all simulations, the magnetic field is growing in the bulk region of the orbital plane, but in the azimuthal direction the magnetic field intensity is noticeably lower.

\subsection{Evolution of the energy contributions}

\begin{figure*}
	\centering
	\includegraphics[width=0.45\linewidth]{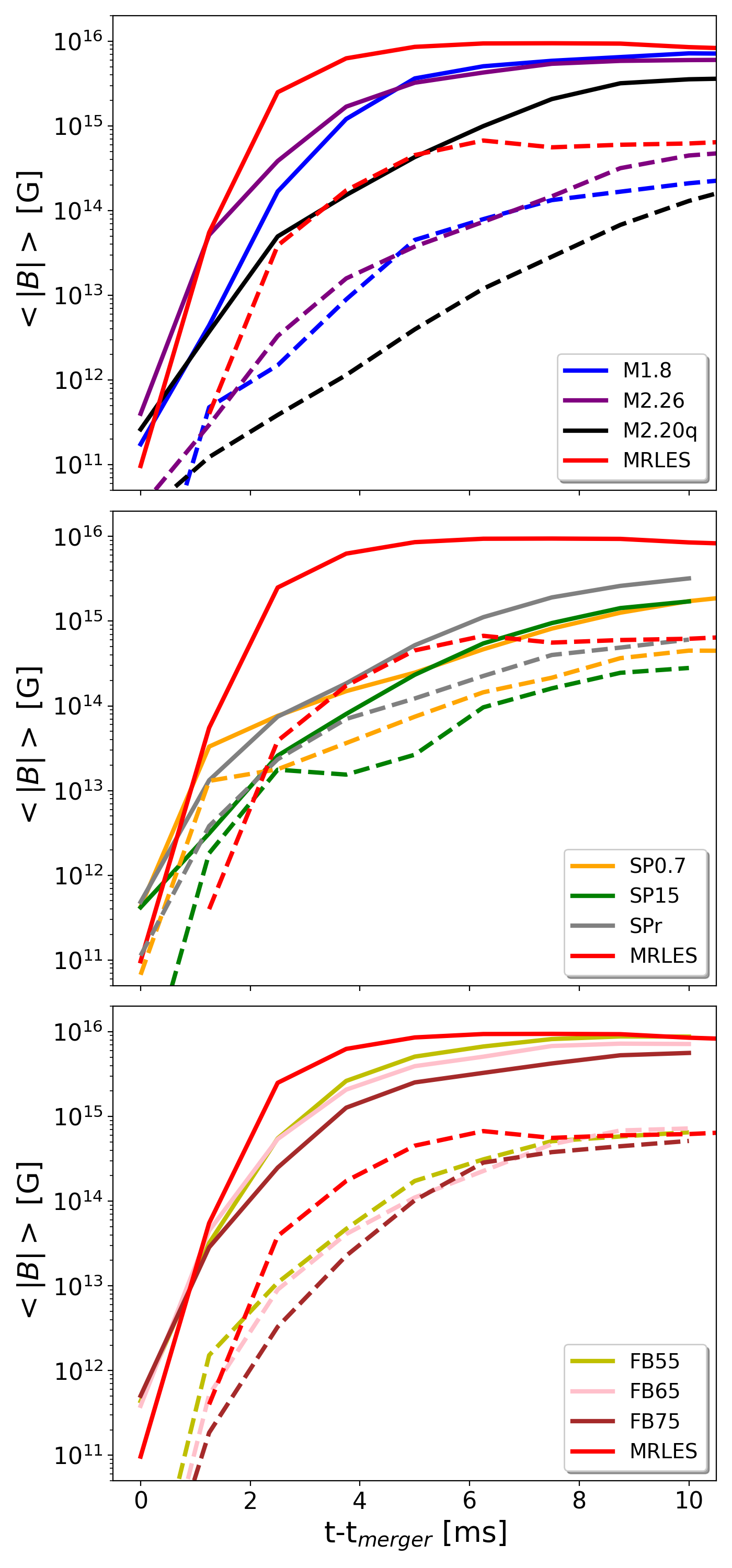}
	\includegraphics[width=0.45\linewidth]{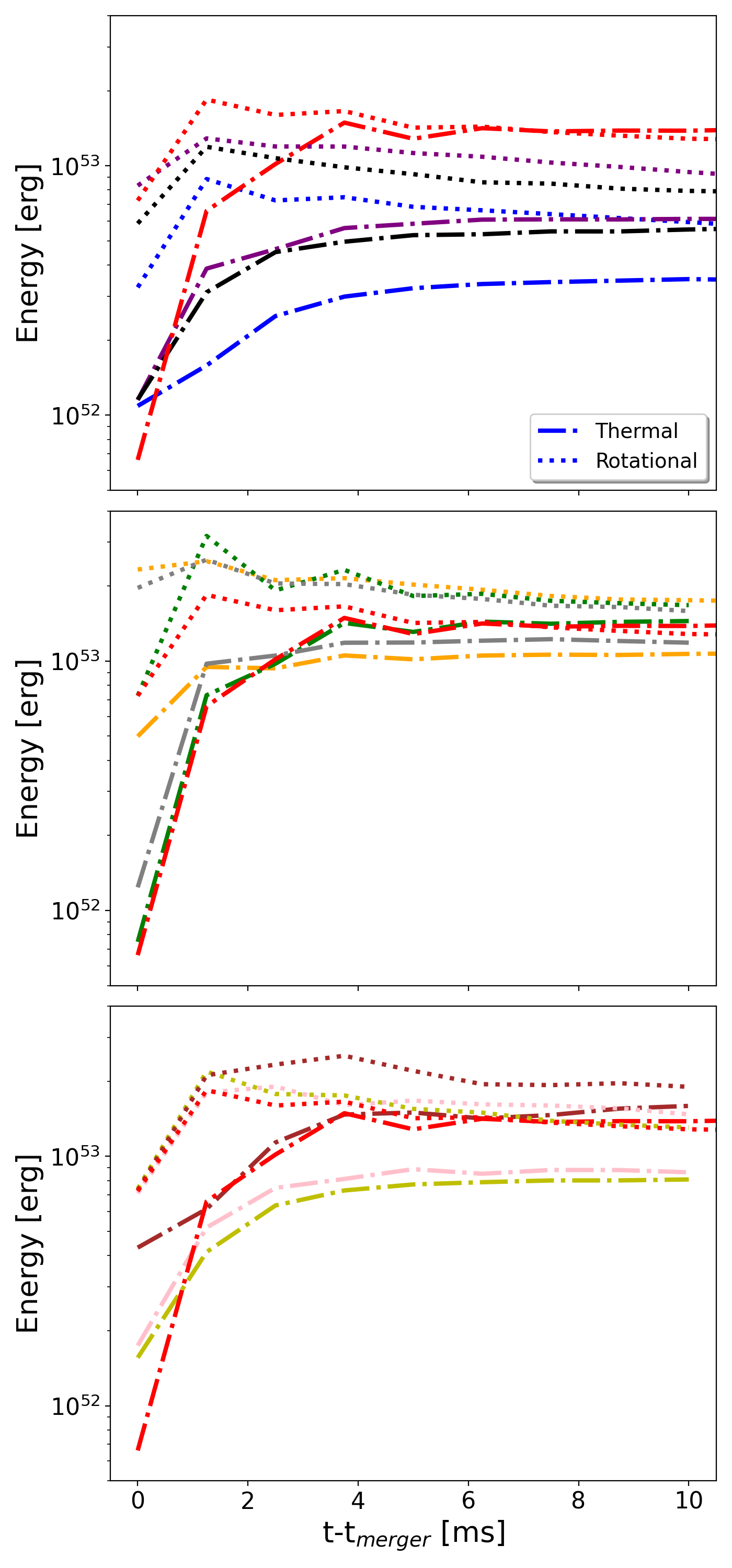}
	\caption{\textit{Volume-averages}. (Left) Evolution of the averaged magnitude of the total magnetic field, both in the bulk (solid lines) and envelope (dashed lines) regimes of the remnant. From top to bottom, it is shown the different variations of mass, mass ratio, spin period and EoS. (Right) Thermal (dotted lines) and rotational (dot-dashed lines) energies as a function of time for the same simulations.}
	\label{fig:volume_average}
\end{figure*}

In Fig.~\ref{fig:volume_average} we show the volume average magnetic field intensity (left) as well as the thermal and rotational energies (right) for a variety of simulations, sampled every $1.25$ ms. We compute separately the magnetic field strength in either the bulk or envelope regions. To reduce cluttering we also separated the different varied quantities (i.e., masses, spin and EoS) into their own rows.

In the first of these rows we show the effect of varying the mass and the mass ratio. Although the growth rate can vary between the different cases, a comparable magnetic field intensity is reached by the end of the simulation (top left panel). This is true for both the bulk and the envelope regions. The thermal and rotational energies for the same simulations span a slightly larger range of values. Both energies clearly rise in systems with larger total masses. Nevertheless, the order of magnitude of the energies does not change.
During the merger, the initial orbital kinetic energy is mostly transformed into rotational energy of the remnant. A significant fraction is instead converted into turbulent kinetic energy, and then transferred to magnetic field energy by means of the small-scale, turbulent dynamo. Notice that there is also a part of the kinetic energy that is converted into thermal energy in the shocked fluid.

For the cases where we vary the stellar spin, the volume-averaged magnetic fields grow slower to large values and have not reached their asymptotic values within the first 10\,ms, especially in the bulk. However, even at this stage they have comparable values of ($2 \cdot 10^{15}$~G) and a straightforward extrapolation suggest that they will reach values of $ \sim 10^{16}$~G a few milliseconds later. The values of the rotational energy of these simulations are even more similar to each other than those with mass variation. Even though a clear ordering is visible, especially in the rotational energy, all simulations reach a thermal energy of about $\sim 10^{53}$~erg. 

Finally, the differences of the magnetic field strengths for the varied EoS simulations are even smaller, and they become negligible after $10$~ms. Despite this, there is a clear ordering from the stiffest EoS at the highest values and the softest EoS at the lowest values. The thermal and rotational energy also converge within a factor of $\sim 2$ between them and are ordered in the opposite direction, with the stiffest EoS being at the bottom.

\subsection{Spectral energy distribution}
\begin{figure*}
	\centering
	\includegraphics[width=\linewidth]{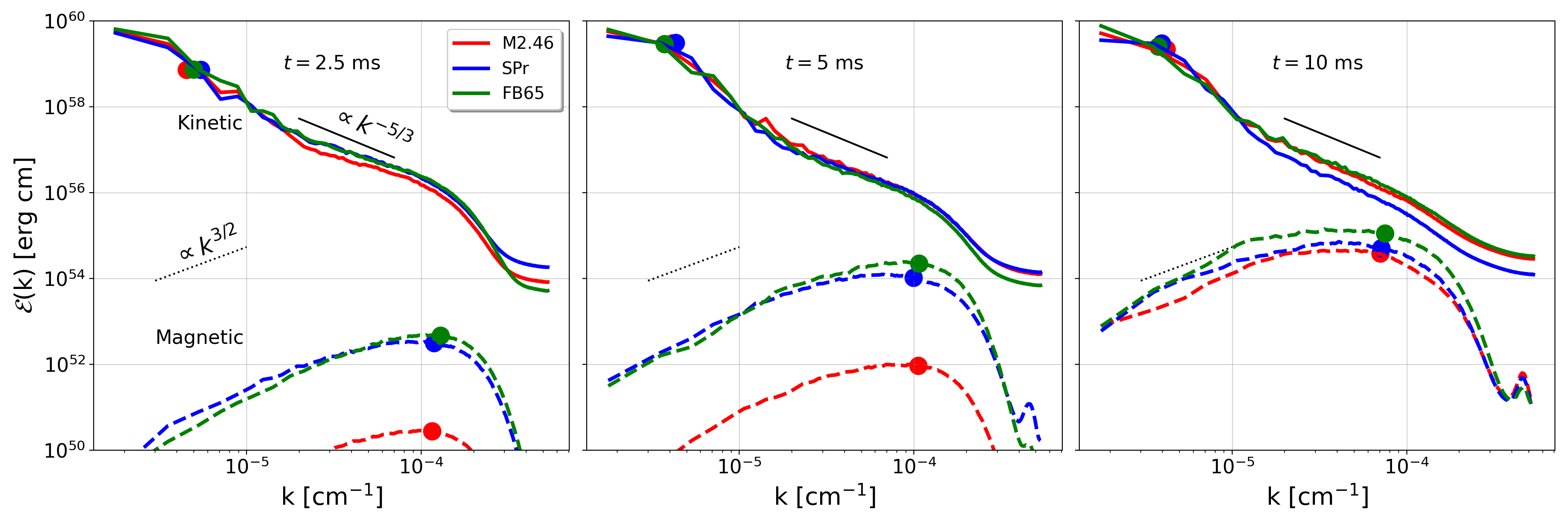}
	\caption{\textit{Energy spectra}. Plots of the magnetic (dashed) and kinetic (solid) energy spectra as a function of the wavenumber at $t=\{2.5,5,10\}$~ms after the merger (first, second and third columns, respectively) for the {\ttfamily M2.46} ($M_{ADM} = 2.46$, $q = 1$, spinningless, EoS$=$ APR4), {\ttfamily SPr} ($M_{ADM} = 2.7$, $q = 1.25$, Spin$_1=1$~ms, Spin$_2 = 10$~ms, EoS$=$ APR4) and {\ttfamily FB65} ($M_{ADM} = 2.7$, $q = 1$, spinningless, EoS$=$ $m/m^* = 0.65$) simulations (red, blue and green colors, respectively). The solid slope corresponds to the Kolmogorov power law and the dotted one to Kazantsev.}
	\label{fig:spectra}
\end{figure*}

The energy spectra for a cube of $18$~km side-length is shown in Fig.~\ref{fig:spectra} for the selected simulations. From left to right, we plot these results at $t=\{2.5, 5, 10\}$~ms after the merger. The solid lines correspond to the kinetic energy, and the dashed lines correspond to the magnetic energy. The solid black slope corresponds to the Kolmogorov's spectra power law ($\propto k^{-5/3})$~\citep{kolmogorov41} and the dotted slope corresponds to the Kazantsev's slope ($\propto k^{3/2})$~\citep{kazantsev68}. Although the simulations start from different initial setups, the shape (but not the magnitude) of the kinetic energy behaves almost identically for each simulation at all represented times. This is because the turbulence transfers kinetic energy from the large scales to the small ones, where the dynamo effect (i.e., the conversion from kinetic to magnetic energy) becomes more effective. The resulting increase of the magnetic energy is similar in every simulation performed, with a difference of a factor of $2-3$ at most. In fact, the magnetic energy has grown during the timespan of the simulations, consistent with the results previously shown in Fig.~\ref{fig:volume_average}.

\subsection{Equipartition scale}

A theoretical estimate of the equipartition scale $\ell_{\rm eq}$, which marks the intersection between the Kolmogorov kinetic cascade and
the Kazantsev magnetic spectrum, might follow from the comparison of two timescales:  the eddy turnover time of the small-scale dynamo $\tau(\ell)$ and the timescale of the magnetic field to reach saturation $t_{\text{sat}}$.

The timescale available for the turbulent dynamo to amplify the magnetic field to saturation in a binary neutron star merger is limited, typically a few rotation periods of the bulk and close envelope of the remnant $ t_{\text{sat}} \sim [5,10] \, \mathrm{ms}$. After that time, the remnant becomes mostly axisymmetric and only a small degree of turbulence remains in the kinetic energy, insufficient to efficiently amplify the magnetic field (see for instance the turbulence indicators \citep{aguilera23}).

The magnetic field starts out weak and is amplified by a turbulent dynamo at small scales and then follows the Kazantsev spectrum at intermediate and large scales. As time goes on, the magnetic energy grows and reaches equipartition with the kinetic energy at progressively larger scales. During this time, the growth of $\ell_{\rm eq}$ via inverse energy transfer from small to larger scales is limited by the saturation time and by causality. According to Kolmogorov`s turbulence theory\citep{kolmogorov41,Landau1987Fluid}, the typical velocity fluctuation at size $\ell$ scales as $v(\ell) \sim (\varepsilon \, \ell)^{1/3}$, being $\varepsilon$ the energy dissipation rate per unit mass. 
The eddy turnover time at such scale $\ell$  must satisfy
\begin{eqnarray}
    \tau(\ell) \sim \frac{\ell}{v(\ell)} \sim \ell^{2/3} \varepsilon^{-1/3}.   
\end{eqnarray}

Since the energy dissipation rate is assumed to be constant across the inertial range, it can be evaluated at the outer scale $L$ (i.e., the largest in the inertial range). By using typical values during the merger, we obtain
\begin{eqnarray}
    \varepsilon \sim 10^{22} \left(\frac{v_L}{0.1 c}\right)^3 \left( \frac{10 \mathrm{km}}{L} 
    \right) \mathrm{cm}^2 \cdot \mathrm{s}^{-3}
\end{eqnarray}

The growth rate of the magnetic field is on the order of the inverse eddy turnover time, while that saturation of the magnetic energy at that scale is reached approximately after $N=20-40$ turnover times \citep{Schekochihin_2004,BRANDENBURG20051}. Requiring that the magnetic field gets amplified and saturates within the available time of strong turbulence, $N \tau(\ell) \lesssim t_{\text{sat}}$, we find that the maximum scale reachable at saturation is $\ell(t) \sim \varepsilon^{1/2} (t_{\text{sat}}/N)^{3/2}$. Using the typical values of our simulations we find the rough estimate range $ \ell_{\rm eq} \sim 1-4\, \mathrm{km}$ for the equipartition scale. The lower end of this estimated range is in reasonable agreement, within the expected uncertainties arising from several simplifications and parameter ambiguities, with the characteristic scale $\langle L \rangle \sim 1\, \mathrm{km}$ found in our spectra analysis.

\section{Conclusions} \label{sec:conclusions}

\begin{figure}
	\centering
	\includegraphics[width=0.9\linewidth]{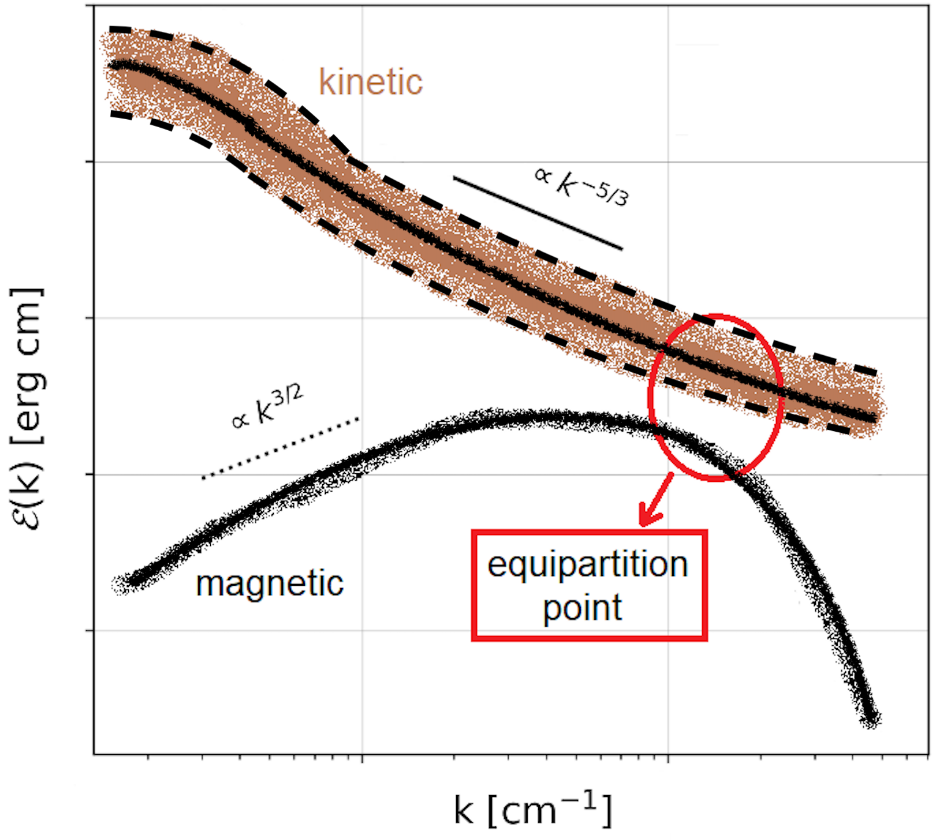}
	\caption{Sketch of the spectra at $t=10$~ms after the merger. The intense brown stroke represents the kinetic energy and the surrounding strokes represent its bounds. The black stroke corresponds to the magnetic energy at $\sim 10$~ms after the merger. Although the the initial conditions of the neutron stars are changed, the shape of the kinetic energy after the merger remains bounded (dashed lines). When the magnetic energy reaches (or gets so close to) the equipartition point (represented here as the red circle) no more kinetic energy is transferred to magnetic energy in the small scales. Then, the magnetic energy can only rise on the intermediate and large scales at later times.}
	\label{fig:spectra_example}
\end{figure}

In this work, we investigated the influence of the total mass, the mass ratio, the spin and the neutron star equation of state on the turbulent amplification of the magnetic field in BNS mergers. Our simulations start with a relatively small initial poloidal large-scale magnetic field of $5 \times 10^{11}$G and cover the first $\sim 10$~milliseconds after the merger. Within this time frame we find that the volume-averaged intensity of the magnetic field, and by extension the magnetic spectra energy, is comparable for all simulations. Specifically, it approaches values of $\sim 10^{16}$~G in the bulk and does not exceed values of $\sim 10^{15}$~G in the envelope regions of the remnant. The differences are within a factor of $\sim 5$. Local values can exceed the volume-averaged values by factors of a few. This result is rather insensitive to the initial conditions. However, the growth rate of the magnetic field intensity is dependent on the initial conditions in a non-trivial way. The close agreement between various systems at $10$~ms after the merger also extends to the thermal and rotational energy, reaching values of $\sim 10^{52}$~erg and $\sim 10^{53}$~erg, respectively.

This behavior is not entirely unexpected. It can be explained by considering that the kinetic energy as a function of the wavelength is fixed at all scales with only little variation. When two neutron stars merge, there is always a shear layer between the stars that becomes Kelvin-Helmholtz unstable and triggers turbulence. The kinetic energy as a function of the wavelength, then, follows the Kolmogorov slope (see the sketch in Fig.~\ref{fig:spectra_example}). Since the kinetic energy is fixed, the amount of energy that can be transferred from it to the magnetic energy is limited by the equipartition point (red circle), which is located at small scales. The equipartion point limits the transfer, because it's the point where the magnetic field saturates and stops growing. Due to a lack of kinetic energy to convert, the magnetic field cannot be increased substantially beyond $\sim 10^{16}$~G in the bulk region, defining a clear limit on the amplification of the magnetic field in this turbulent phase. A similar effect was observed in \citet{aguilera22}, where the initial topology and intensity of the magnetic field were varied. This has implications for the ability to launch a jet. The limit of $10^{16}\,$G found here can only be surpassed by assuming an unrealistically high large-scale poloidal magnetic field strength of $B_{\mathrm{0}} \ge 10^{15}\,$G just before the merger~\citep{aguilera24}. Therefore, we conclude that the magnetic field amplification in a realistic scenario is not strongly correlated with initial parameters of the system, like mass, mass ratio, stellar spin and EoS.

To summarize: Even though it is possible to find a potentially infinite number of possible initial configurations for neutron stars in a merger, we find that large differences in those configurations yield comparable results. We therefore strongly suggest that the magnetic field amplification at $10$~ms after a merger would look very similar to the results presented above, and are insensitive to detailed neutron star binary properties. Note, however, that this statement only concerns the magnetic field amplification and not other quantities that might be effected by the initial conditions. We also have not yet investigated more exotic scenarios, like quark or dark matter, which we plan to investigate in a future work. However, we do not expect our results to change significantly, as the Kolmogorov slope should be unaffected by such exotic scenarios.

\subsection*{Acknowledgments}

The authors thank Thomas Tauris for insightful discussions about binary stellar evolution and the NS-NS parameter space. RA-M is funded by the Deutsche Forschungsgemeinschaft (DFG, German Research Foundation) under the Germany Excellence Strategy - EXC 2121 'Quantum Universe' - 390833306. JEC and SR are funded by the European Research Council (ERC) Advanced Grant INSPIRATION under the European Union’s Horizon 2020 research and innovation program (Grant agreement No. 101053985). SR has been supported by the Swedish Research Council (VR) under grant number 2020-05044, by the research environment grant 'Gravitational Radiation and Electromagnetic Astrophysical Transients' (GREAT) funded by the Swedish Research Council (VR) under Dnr 2016-06012, by the Knut and Alice Wallenberg Foundation under grant Dnr. KAW 2019.0112, by Deutsche Forschungsgemeinschaft (DFG, German Research Foundation) under Germany’s Excellence Strategy - EXC 2121 'Quantum Universe' - 390833306. This work was supported by the Grant PID2022-138963NB-I00 funded by MCIN/AEI/10.13039/501100011033/FEDER, UE. The authors gratefully acknowledge the computing time made available to them on the high-performance computer "Lise" at the NHR Center NHR@ZIB. This center is jointly supported by the Federal Ministry of Education and Research and the state governments participating in the NHR~\citep{nhr}.

\subsection*{Data availability}

The data generated is available from the correspond author upon reasonable request.


\bibliographystyle{mnras}
\bibliography{turbulence} 



\bsp	
\label{lastpage}
\end{document}